\magnification=1200
\hfuzz=3pt
\hsize=12.5cm
\hoffset=0.32cm
\baselineskip=18pt
\voffset=\baselineskip
\nopagenumbers
\font\title=cmssdc10 at 20pt
\font\smalltitle=cmssdc10 at 14pt

\let\text=\textstyle

\def\RE{\mathop{\Re e}\nolimits}

\def\sectionstyle{\smalltitle}
\newskip\beforesectionskip
\newskip\aftersectionskip
\beforesectionskip=4mm plus 1mm minus 1mm
\aftersectionskip=2mm plus .2mm minus .2mm
\newcount\mysectioncounter
\def\resetsections{\mysectioncounter=0}
\resetsections
\newcount\myeqcounter
\def\mysection#1\par{\par\removelastskip\penalty -250
\vskip\beforesectionskip
\global\advance\mysectioncounter by 1\noindent
\myeqcounter=0{\sectionstyle\the\mysectioncounter.
#1}\par
\nobreak\vskip\aftersectionskip}
\def\myeqno{\global\advance\myeqcounter by 1\eqno{(\the\mysectioncounter.\the\myeqcounter)}}
\def\mydispeqno{\global\advance\myeqcounter by 1\hfill\llap{(\the\mysectioncounter
.\the\myeqcounter)}}
\headline={\hfil\tenrm\folio\hfil}
\footline={\hfil}
\baselineskip=12pt
\parindent= 20pt
\centerline{\smalltitle Aging Effects in Free Quantum Brownian Motion}
\vskip 0.5cm
\centerline{No\"elle POTTIER}  
\centerline{\sl Groupe de Physique des Solides\footnote{$^1$}{\rm Laboratoire
associ\'e au C.N.R.S. (U.M.R. n$^0\ 7588$) et aux Universit\'es Paris 7 et Paris
6.}, Universit\'e Paris 7,}
\centerline{\sl Tour 23, 2 place Jussieu, 75251 Paris Cedex 05, France,\/}
\smallskip
\centerline{and}
\smallskip
\centerline{Alain MAUGER}
\centerline{\sl Laboratoire des Milieux D\'esordonn\'es et
H\'et\'erog\`enes\footnote{$^2$}{\rm Laboratoire associ\'e au C.N.R.S. (U.M.R. n$^0\ 7603$) et \`a
l'Universit\'e Paris 6.},}
\centerline{\sl Universit\'e Paris 6, Tour 22, Case 86,}
\centerline{\sl 4 place Jussieu, 75252 Paris Cedex 05, France.\/}
\vskip 0.5cm
\noindent
{\smalltitle Abstract}
\medskip
\noindent
The two-time correlation function $C_{xx}(t,t')$ of the displacement $x(t)-x(t_0)$ of a free quantum
Brownian particle with respect to its position at a given time~$t_0$ is calculated analytically in
the framework of the Caldeira and Leggett ohmic dissipation model. As a result, at any temperature
$T$, $C_{xx}(t,t')$ exhibits aging, {\sl i.e.\/} it depends explicitly on both times $t$ and $t'$
and not only on the time difference $\tau=t-t'$, even in the limit of large age $t'$
($t_0\leq t'\leq t$), in contrast with a dynamic variable in equilibrium such as the particle
velocity. The equilibrium quantum fluctuation-dissipation theorem (QFDT) has to be modified in
order to relate the response function $\chi_{xx}(t,t')$ to $C_{xx}(t,t')$, since this latter
quantity takes into account even those fluctuations of the displacement which take place during the
waiting time $t_w=t'-t_0$. We describe the deviation from QFDT in terms of an effective inverse
temperature $\beta_{\rm eff.}(\tau,t_w)$. The behavior of this quantity as a function of $\tau$ for
given values of $T$ and $t_w$ is analyzed. In the classical limit it is shown that $\beta_{\rm
eff.}(\tau,t_w)=\beta {D(\tau)/[D(\tau)+D(t_w)]}$, where $D(t)$ denotes the time-dependent diffusion
coefficient.
\medskip
\parindent=0pt
{\bf PACS numbers:} 

05.30.-d Quantum statistical mechanics.

05.40.+j Fluctuation phenomena, random processes, and Brownian motion.

02.50.Ey Stochastic processes
\medskip
\medskip
{\sl Corresponding author:

No\"elle POTTIER

Groupe de Physique des Solides, Universit\'e Paris 7,

Tour 23, 2 place Jussieu, 75251 Paris Cedex 05, France

Fax number: +33 1 43 54 28 78. E-mail: pottier@gps.jussieu.fr\/}
\vfill
\break
\parindent=20pt
\baselineskip=12pt

\mysection{Introduction}

Out-of-equilibrium dynamics, characterized by slow relaxation and aging effects, is currently the
subject of numerous studies. Aging is by definition the property that the correlation
function $\langle O(t)O(t')\rangle$ of an observable $O$ at two subsequent times $t'$ and $t$
($t'\leq t$) is not invariant under time translation even in the limit of large age $t'$ of the
system, in contrast with a dynamic variable in equilibrium. In other words, this correlation
function depends both on the time difference
$\tau=t-t'$ and on the waiting time (or age) $t_w=t'$. In such a non-equilibrium situation,
the fluctuation-dissipation theorem (FDT) is violated. The aging phenomenon has been intensively
studied, both  experimentally and theoretically, mostly in complex systems such as
spin-glasses and other types of glasses~[1]. 

Interestingly enough, aging can also be encountered in simpler, non disordered systems [2]-[6]. For
instance, as first noticed in [2], the simplest example of a dynamical variable which does not reach
equilibrium and exhibits aging is the random walk: indeed the correlation function of the
displacement of a free Brownian particle at two different times depends explicitly on the two
times involved, and not simply on their difference. In this case, the violation of the
fluctuation-dissipation theorem can simply be characterized by a constant factor rescaling the
temperature [2].

Up to now, aging effects have mostly been discussed in a classical framework (quantum aging
in quantum spin systems have however been studied in [7]-[8]). In order to investigate quantum aging
effects, it is interesting, to begin with, to study the displacement of a free
quantum Brownian particle, since, as quoted above, this variable exhibits aging in the classical
case. Free quantum Brownian motion can conveniently be described in the framework of the
Caldeira and Leggett ohmic dissipation model [9]-[13]. The dissipation is introduced {\sl via\/} a
linear coupling of the particle to a thermal bath with a continuum of modes. In the ohmic model,
this coupling is proportional to $\omega$ at low frequency, an assumption which allows, under
certain conditions, to recover an equation of motion of the Langevin type. However, up to now, only
equal time correlation functions have been computed in this framework and the question of aging has
not yet been considered.
 
The study of correlation functions at low temperature is all the more interesting since the
interplay between quantum and thermal fluctuations then becomes effective. For instance, at low
temperature, the equal time velocity correlation function displays a long time tail varying like
$-\,t^{-2}$, which leads to a non-classical behaviour of the time-dependent diffusion coefficient
$D(t)$ as a function of time~[12]. In the same way, non-classical effects may also be expected to
take place in the correlation functions at two different times. This study is the aim of the present
paper in which we compute analytically in an exact manner the two-time velocity and displacement
correlation functions of a free quantum Brownian particle in the framework of the ohmic
dissipation model. As a result, the two-time displacement correlation function displays aging
whatever the temperature of the bath. The model allows for a discussion of the modifications of the
corresponding quantum fluctuation-dissipation theorem (QFDT) due to aging.

The paper is organized as follows. In section 2, we recall briefly the results obtained in [2] for
the two-time displacement correlation function in the viscous limit of the classical Langevin model
and the way the classical FDT has to be modified. This modification can be viewed as a simple
rescaling of the temperature, the inverse temperature $\beta={1/k_BT}$ being replaced by an
effective inverse temperature $\beta_{\rm eff.}={\beta/2}$. In section 3, remaining in the
classical framework, we take the particle inertia into account. Then, the FDT violation may still be
described by a rescaling of the temperature. The effective inverse temperature now effectively
depends of the two times $\tau$ and $t_w$ : $\beta_{\rm eff.}=\beta_{\rm eff.}(\tau,t_w)$.
Interestingly enough, this dependence takes place {\sl via\/} one-time quantities, namely the
time-dependent diffusion coefficients $D(\tau)$ and $D(t_w)$:
$\beta_{\rm eff.}(\tau,t_w)=\beta{D(\tau)/[D(\tau)+D(t_w)]}$. In section~4, turning to the
description of quantum Brownian motion, we briefly recall the main features of the Caldeira and
Leggett ohmic dissipation model [9]-[13]. Then, in section 5, we compute analytically the two-time
velocity and displacement correlation functions of a free quantum Brownian particle. We analyze in
detail the behaviour of the quantum time-dependent diffusion coefficient $D^Q(t)$ as a function of
time. This behaviour depends on the bath temperature in a way which allows a clear-cut distinction
between a classical and a quantum regime for $D^Q(t)$. In section 6, we discuss, in the specific
case of the diffusion of a free quantum Brownian particle, the way the quantum
fluctuation-dissipation theorem (QFDT) has to be modified whenever aging is present. We show how
the deviations from QFDT can, for any temperature of the bath, be described in terms of an
associated effective inverse temperature
$\beta^Q_{\rm eff.}(\tau,t_w)$, which can be computed from the quantum time-dependent diffusion
coefficients
$D^Q(\tau)$ and $D^Q(t_w)$. However, the QFDT violation cannot, except in the classical
limit, be reduced to a simple rescaling of the bath temperature. The behavior of $\beta^Q_{\rm
eff.}(\tau,t_w)$ as a function of $\tau$ for given values of $T$ and $t_w$ is analyzed. In
Section 7, aging effects in the quantum diffusion problem are discussed. Finally, Section 8 contains
our conclusions.

\mysection{Aging effects in classical Brownian motion : the viscous limit of the Langevin model}

In the viscous limit of the Langevin model, the one-dimensional dynamics of the
free Brownian particle of viscosity $\eta=m\gamma$ is described by the first-order Langevin equation
$$m\gamma\,{dx\over dt}=F(t),\myeqno$$
in which the Langevin force $F(t)$ is modelized by a stationary random Gaussian process of zero mean
and of correlation function
$$\langle F(t)F(t')\rangle=2k_BTm\gamma\,\delta(t-t'),\myeqno$$
where $T$ is the temperature. (The symbol $\langle\ldots\rangle$ denotes the average value over the
realizations of the noise).

Let us choose some time $t_0$ and consider the displacement of the particle at any later time $t\geq
t_0$, as defined by
$$x(t)-x(t_0)={1\over m\gamma}\int_{t_0}^tdt_1\,F(t_1).\myeqno$$
\bigskip
{\bf 2.1. The two-time displacement correlation function and the displacement response function}

Let us consider the two-time displacement correlation function $C_{xx}(t,t';t_0)$ as defined by:
$$C_{xx}(t,t';t_0)=\langle[x(t)-x(t_0)][x(t')-x(t_0)]\rangle.\myeqno$$
It is easy to check that, for $t_0\leq t'\leq t$,
$$C_{xx}(t,t';t_0)=2{k_BT\over m\gamma}\,(t'-t_0),\myeqno$$
or, in terms of the time difference $\tau=t-t'$ and of the waiting time $t_w=t'-t_0$,
$$C_{xx}(\tau,t_w)=2{k_BT\over m\gamma}\,t_w.\myeqno$$
One recovers the fact that in this model the two-time displacement correlation function is
proportional to the waiting time [2], [15]. 

The displacement correlation function $C_{xx}(\tau,t_w)$ is not simply a function of the time
difference $\tau$ (in this very simple model, it even does not depend on $\tau$). In other words,
the displacement $x(t)-x(t_0)$ of the free Brownian particle is not a stationary random process.
Thus, Fourier analysis and the Wiener-Khintchine theorem are not applicable to this quantity.

As for the displacement response function $\chi_{xx}(t,t')$, by definition, it characterizes the
average displacement $\langle x(t)-x(t_0)\rangle$ at time $t$ due a unit impulse of force taking
place at a previous time
$t'<t$. This response function is easily deduced from the equation of motion (2.1) in which one
adds to the Langevin force a non random force proportional to $\delta(t-t')$. One gets 
$$\chi_{xx}(t,t')=\Theta(t-t')\,{1\over m\gamma},\myeqno$$ 
where $\Theta(t-t')$ denotes the unit step function.
\bigskip
{\bf 2.2. The modified fluctuation-dissipation theorem}

As first noticed in [2], the fluctuation-dissipation theorem (FDT) has to be generalized in order to
handle such a case. The relation between a response function and the associated correlation
function,
$$\chi(t,t')=\beta\,\Theta(t-t')\,{\partial C(t,t')\over\partial t'}\qquad(t'< t),\myeqno$$
valid for a dynamic variable in equilibrium, is clearly not applicable to the Brownian particle
displacement $x(t)-x(t_0)$. As proposed in [2], one can write the actual relation between
$\chi_{xx}(t,t')$ and $C_{xx}(t,t';t_0)$ with $t_0\leq t'\leq t$ as
$$\chi_{xx}(t,t')=\beta\,\Theta(t-t')\,X(t,t';t_0)\,{\partial C_{xx}(t,t';t_0)\over\partial
t'}\qquad(t_0\leq t'< t),\myeqno$$  
where the quantity $X(t,t';t_0)$ acts as a violation factor of the
fluctuation-dissipation theorem. Defining for the dynamic variable under study an effective
temperature $T_{\rm eff.}(t,t';t_0)={1/k_B\beta_{\rm eff.}(t,t';t_0)}$ by
$$\beta_{\rm eff.}(t,t';t_0)=\beta\,X(t,t';t_0),\myeqno$$
one can rewrite Eq. (2.9) as
$$\chi_{xx}(t,t')=\beta_{\rm eff.}(t,t';t_0)\,\Theta(t-t') {\partial C_{xx}(t,t';t_0)\over\partial
t'}\qquad(t_0\leq t'< t).\myeqno$$  

In the viscous limit of the Langevin model, using the expressions (2.5) and (2.7) of
$C_{xx}(t,t';t_0)$ and
$\chi_{xx}(t,t')$, one gets, for any $t$ and $t'$ ($t'<t$):
$$X(t,t';t_0)={1\over 2}.\myeqno$$
Thus, in this simple description of classical Brownian motion, the FDT violation factor is a
constant [2]. Moreover, since it does not depend on $T$, it can simply be viewed as rescaling the
temperature, as far as the variable of interest is concerned~:
$$\beta_{\rm eff.}={\beta\over 2}.\myeqno$$ 

\mysection{Aging effects in classical Brownian motion : the Langevin model}

Actually, the full non retarded Langevin equation contains an inertial term and writes down
$$m\,{dv\over dt}+m\gamma\,v=F(t),\qquad v={dx\over dt},\myeqno$$
where $F(t)$ is the above defined Gaussian delta-correlated random force.
\bigskip
{\bf 3.1. The velocity correlation function}

If, at a given initial time $t_i$ the velocity takes the value $v_i$, at any
later time $t\geq t_i$ it is given by
$$v(t)=v_i\,e^{-\gamma(t-t_i)}+{1\over m}\int_{t_i}^t dt_1\,F(t_1)\,e^{-\gamma(t-t_1)}.\myeqno$$

The velocity correlation function, as defined by
$$C_{vv}(t_1,t_2)=\langle v(t_1)v(t_2)\rangle,\myeqno$$
is equal to
$$C_{vv}(t_1,t_2)=(v_i^2-{k_BT\over m})\,e^{-\gamma(t_1+t_2-2t_i)}+{k_BT\over
m}e^{-\gamma|t_1-t_2|}.\myeqno$$

Letting $t_i\to -\infty$, the velocity $v(t)$ reduces to its stationary part
$$v(t)={1\over m}\int_{-\infty}^t dt_1\,F(t_1)\,e^{-\gamma(t-t_1)}\myeqno$$ 
and, correspondingly, the correlation function $C_{vv}(t_1,t_2)$ becomes the equilibrium correlation
function
$$C_{vv}(t_1-t_2)={k_BT\over m}e^{-\gamma|t_1-t_2|},\myeqno$$
which only depends on the time difference $t_1-t_2$. Note that this latter result
can also be obtained by taking as initial condition for the velocity a thermally distributed random
variable $v_i$.

The velocity, which equilibrates at large times, is not an aging variable. Thus, Fourier
analysis and the Wiener-Khintchine theorem can be used equivalently in order to obtain the
equilibrium velocity correlation function $C_{vv}(t_1-t_2)$. 
\bigskip
{\bf 3.2. The two-time displacement correlation function and the displacement response function} 

From now on, we assume that the limit $t_i\to -\infty$ has been taken. The displacement
correlation function $C_{xx}(t,t';t_0)$ as defined by Eq. (2.4) can be obtained from the
equilibrium velocity correlation function by computing the double integral
$$C_{xx}(t,t';t_0)=\int_{t_0}^t dt_1\int_{t_0}^{t'}dt_2\,C_{vv}(t_1-t_2).\myeqno$$
Using Eq. (3.6), one gets, for $t_0\leq t'\leq t$,
$$C_{xx}(t,t';t_0)={k_BT\over
m\gamma}\,\left[2(t'-t_0)-{1+e^{-\gamma(t-t')}-e^{-\gamma(t-t_0)}-
e^{-\gamma(t'-t_0)}
\over\gamma}\right],\myeqno$$
or, in terms of the time difference $\tau=t-t'$ and of the waiting time $t_w=t'-t_0$,
$$C_{xx}(\tau,t_w)={k_BT\over
m\gamma}\,\left[2t_w-{1+e^{-\gamma\tau}-e^{-\gamma(\tau+t_w)}-e^{-\gamma
t_w}\over\gamma}\right].\myeqno$$
Thus, as in the viscous limit, the displacement $x(t)-x(t_0)$ of the free Brownian particle is not
a stationary random process.

As for the displacement response function $\chi_{xx}(t,t')$, when one adds to the Langevin force in
Eq. (3.1) a non random force proportional to $\delta(t-t')$, one gets:
$$\chi_{xx}(t,t')=\Theta(t-t')\,{1\over
m\gamma}\,[1-e^{-\gamma(t-t')}].\myeqno$$
This quantity only depends on the time difference $\tau=t-t'$.
\bigskip
{\bf 3.3. The FDT violation factor and the effective temperature in the Langevin model}

According to Eqs. (3.8) and (3.9), the FDT violation factor in Eq. (2.9) is given by
$$X(t,t';t_0)={1-e^{-\gamma(t-t')}\over 2-e^{-\gamma(t-t')}-e^{-\gamma(t'-t_0)}},\myeqno$$
or, in terms of $\tau$ and $t_w$,
$$X(\tau,t_w)={1-e^{-\gamma\tau}\over 2-e^{-\gamma\tau}-e^{-\gamma t_w}}.\myeqno$$
Since $X(\tau,t_w)$ does not depend on $T$, it can still be viewed as rescaling the
temperature.  The rescaling factor now depends on both times $\tau$ and $t_w$. In the limit of
large $\tau$ and $t_w$ ($\gamma\tau\gg 1$, $\gamma t_w\gg 1$), it tends towards the constant value
${1/2}$ which corresponds to the viscous limit of the Langevin model (Eq. (2.12)). As for the
associated effective inverse temperature, since one has
$$\beta_{\rm eff.}(\tau,t_w)=\beta\,X(\tau,t_w),\myeqno$$
the discussion of the behaviour of this quantity as a function of $\tau$ and $t_w$ can be
reduced to that for the FDT violation factor.  
\bigskip
{\bf 3.4. Analysis of the violation factor and of the effective temperature in terms of the
time-dependent diffusion coefficient}

In order to analyze at best the behaviour of $X(\tau,t_w)$ and of $\beta_{\rm eff.}(\tau,t_w)$ as a
function of the two arguments $\tau$ and $t_w$, it is convenient to introduce the time-dependent
diffusion coefficient $D(t)$, which is an odd function of time defined by
$$D(t)={1\over 2}\,{d\over dt}\langle[x(t)-x(0)]^2\rangle.\myeqno$$
One can write
$$D(t)={1\over 2}\,{d\over dt}\,C_{xx}(t,t;0).\myeqno$$
For $t>0$, one can use in Eq. (3.15) the above derived expression for $C_{xx}(t,t;0)$. At any
time $t$, one can write 
$$D(t)=\int_0^t du\,C_{vv}(u).\myeqno$$

In the Langevin model (3.1), using the expression (3.6) of the equilibrium velocity correlation
function, one gets:
$$D(t)={k_BT\over m\gamma}\,\bigl(1-e^{-\gamma t}\bigr),\qquad t>0.\myeqno$$
For $t>0$, $D(t)$ increases monotonously with $t$. In the limit $\gamma t\gg 1$, it tends towards
the usual Einstein diffusion coefficient $D={k_BT/m\gamma}$.

The FDT violation factor (3.12) can be rewritten in terms of the
time-dependent diffusion coefficients $D(\tau)$ and $D(t_w)$ as
$$X(\tau,t_w)={D(\tau)\over D(\tau)+D(t_w)}.\myeqno$$
This quantity is obviously smaller than 1 and, as in the viscous limit, does not depend on $T$.
The associated effective inverse temperature is given by
$$\beta_{\rm eff.}(\tau,t_w)=\beta\,{D(\tau)\over D(\tau)+D(t_w)}.\myeqno$$
Interestingly enough, $X$ and $\beta_{\rm eff.}$ depend on $\tau$ and $t_w$
through the corresponding time-dependent diffusion coefficients\footnote{$^1$}{In fact, the
validity of Eqs. (3.18) and (3.19) is not restricted to the simple ({\sl i.e.\/} non retarded)
Langevin model as described by Eq. (3.1). These formulas can be applied in other classical
descriptions of Brownian motion in which a time-dependent diffusion coefficient can be defined, for
instance when the motion of the free Brownian particle is described by a classical retarded
Langevin equation.}
$D(\tau)$ and $D(t_w)$. 

Actually, formula (3.18) can be understood as follows. Coming back to the definition (2.9) of
$X(t,t';t_0)$, one sees that the numerator of this quantity is proportional to the response function
$\chi_{xx}(t,t')$ while its denominator is proportional to the derivative ${\partial
C_{xx}(t,t';t_0)/\partial t'}$. Between $t_0$ and $t'$ ({\sl i.e.\/} during the waiting time
$t_w$), the diffusing particle does not move in the average: $\langle x(t')\rangle=\langle
x(t_0)\rangle$. Clearly, in order to compute the response of the displacement to an impulse of
force at time $t'$, one cannot apply the fluctuation-dissipation theorem in its
standard form during the full time interval $(t_0,t)$ ({\sl i.e.\/} to $x(t)-x(t_0)$), but only
during the restricted time interval $(t',t)$ ({\sl i.e.\/} to $x(t)-x(t')$) since in this latter
case there is no waiting time. Indeed, doing so,  that is writing
$$\chi_{xx}(t,t')=\beta\,\Theta(t-t')\,{\partial C_{xx}(t,t';t_0)\over\partial
t'}\biggr|_{t_0=t'},\myeqno$$ 
one correctly gets for the displacement response function the expression
$$\chi_{xx}(t,t')=\beta\,\Theta(t-t')\,\int_{t'}^t dt_1\langle v(t_1)v(t')\rangle,\myeqno$$ 
or
$$\chi_{xx}(t,t')=\beta\,\Theta(t-t')\,D(t-t').\myeqno$$
In the Langevin model (3.1), in which the time-dependent diffusion coefficient is given by Eq.
(3.17), the formula (3.22) for $\chi_{xx}(t,t')$ is identical to the expression (3.10), as it
should. As for the denominator of
$X(t,t';t_0)$, it involves the derivative 
$${\partial C_{xx}(t,t';t_0)\over\partial t'}=\int_{-t_w}^\tau du\,C_{vv}(u),\myeqno$$
that is, since $C_{vv}(u)$ is an even function of $u$ (Eq. (3.6)),
$${\partial C_{xx}(t,t';t_0)\over\partial t'}=D(\tau)+D(t_w).\myeqno$$
This result can be easily understood in view of the fact that fluctuations
in the particle displacement $x(t)-x(t_0)$ do take place even during the waiting time $t_w=t'-t_0$.

Summing up, the violation factor $X(t,t';t_0)$ and the associated effective inverse temperature
$\beta_{\rm eff.}(t,t';t_0)$ allow to write a modified fluctuation-dissipation
theorem relating the response function $\chi_{xx}(t,t')$ to the derivative of a correlation function
${\partial C_{xx}(t,t';t_0)/\partial t'}$ with $t_0\leq t'<t$, the latter taking into account even
those fluctuations of the particle displacement which take place during the waiting time. 

Note however that the parameter $T_{\rm eff.}={1/\beta_{\rm eff.}}$, if it allows to
write a modified FDT for the displacement of the free Brownian particle (Eq.~(2.11)), cannot be
given the full significance of a physical temperature. In particular, it does not control the
thermalization of the free Brownian particle. Indeed, in the absence of potential, the
thermalization, which is solely linked with the behaviour of the particle velocity, is
effective once the limit $t_i\to -\infty$ has been taken. 
\bigskip
{\bf 3.5. Discussion of the behaviour of the violation factor and of the effective temperature
in the Langevin model}

Let us now come back to the expression (3.12) of $X(\tau,t_w)$ in the classical non retarded
Langevin model (Eq. (3.1)). Clearly, for all values of $\tau$, $X$ is equal to 1 when $t_w=0$
and it differs from 1 when $t_w\neq 0$. In that sense, it can be said that aging is always
present as far as the displacement of the free Brownian particle is concerned. The curves
representing $X(\tau,t_w)={\beta_{\rm eff.}(\tau,t_w)/\beta}$ as a function of the time difference
$\tau$ for various non zero values of the waiting time $t_w$ are plotted on Fig.~1. For a given
value of $t_w$, $X(\tau, t_w)$ and $\beta_{\rm eff.}(\tau,t_w)$ -- like $D(\tau)$ -- increase
monotonously with $\tau$ towards a finite limit $X_\infty(t_w)$.

Let us discuss the limit value of $X$ at large values of $\tau$ ($\gamma\tau\gg 1$). In this limit,
$D(\tau)$ as given by Eq. (3.17) with $t$ replaced by $\tau$ is equal to the Einstein diffusion
coefficient $D={k_BT/m\gamma}$. For a given $t_w$, the limit value of $X$ at large values of $\tau$
is thus 
$$X_\infty(t_w)\equiv\lim_{\gamma\tau\gg 1}X(\tau,t_w)={1\over 2-e^{-\gamma
t_w}}.\myeqno$$ For small values of $t_w$ ($\gamma t_w\ll 1$), $X_\infty(t_w)$ does not depart very
much from 1. For large values of $t_w$ ($\gamma t_w\gg1$), $X_\infty(t_w)$ approaches the value
${1/2}$. The result (3.25) interpolates in the diffusive regime between the situation with no
waiting time and the situation with an infinite waiting time. 
 
In the opposite limit, {\sl i.e.\/} for small values of $\tau$ ($\gamma\tau\ll 1$), the motion of
the particle is ballistic, and one gets
$$X\simeq{\gamma\tau\over 1-e^{-\gamma t_w}}.\myeqno$$
For small values of $t_w$ ($\gamma t_w\ll 1$), one has, in this regime,
$$X\simeq{\tau\over\tau+t_w},\myeqno$$
and, for large values of $t_w$,
$$X\simeq\gamma\tau.\myeqno$$

Finally, let us note that, recovering as usual the viscous limit of the Langevin model by letting
$m\to 0$, $\gamma\to\infty$ with the viscosity $\eta=m\gamma$ remaining constant, the factor
$X(\tau,t_w)$ as given by Eq. (3.12) indeed reduces to the constant value ${1/2}$ (Eq. (2.12)) for
any $\tau$, as it should. 

\mysection{The ohmic dissipation model}

Let us now briefly recall the main features of the ohmic dissipation model, which allows for a
proper description of quantum Brownian motion [9]-[13].
\bigskip
{\bf 4.1. The Caldeira and Leggett model}

In quantum mechanics, the equation of motion of a Brownian particle can be obtained by
coupling it to a thermal bath constituted by a set of harmonic oscillators. In the
Caldeira and Leggett model [11], and for a free Brownian particle ({\sl i.e.\/} in the absence of an
external potential, which is the situation considered here), the Hamiltonian of the
particle-plus-bath system reads, in obvious notations,
$$H={p^2\over 2m}-x\sum_\nu\lambda_\nu(b_\nu+b_\nu^\dagger)+\sum_\nu\hbar\omega_\nu
b_\nu^\dagger b_\nu+ x^2\sum_\nu{\lambda_\nu^2\over\hbar\omega_\nu},\myeqno$$
where $\lambda_\nu$ is a real coupling constant and the last counterterm insures the translational
symmetry. From this Hamiltonian, the equations of motion for all the degrees of freedom can be
readily written down. The bath variables once eliminated, the Heisenberg particle position
operator $x(t)$ is seen to obey the retarded equation of motion  
$$m\ddot{x}(t)+m\int_{t_i}^t dt'\,k(t-t')\,\dot{x}(t')=-m\,x(t_i)\,k(t-t_i)+F(t),\myeqno$$
where $t_i$ denotes the initial time at which the interaction between the particle and the bath is
switched on [11]-[12].  In Eq. (4.2), $F(t)$ acts as a random force and $k(t)$ as a retarded memory
kernel. These quantities have the microscopic expressions
$$k(t)=\sum_\nu{2\lambda_\nu^2\over m\hbar\omega_\nu}\,\cos\omega_\nu t,\myeqno$$
and
$$F(t)=\sum_\nu\lambda_\nu[b_\nu(t_i) e^{-i\omega_\nu(t-t_i)}+b_\nu^\dagger(t_i) e^{i\omega_\nu
(t-t_i)}].\myeqno$$
\bigskip
{\bf 4.2. The ohmic model}

In order to induce a relaxation, the oscillators of the bath must be spread over a continuum of
frequencies. A central ingredient in the continuum limit of the model is the product of the density
of modes of the bath, $g(\omega)$, times the squared coupling strength, $|\lambda(\omega)|^2$. One
sets:
$$\Lambda(\omega)=2\pi\,g(\omega)|\lambda(\omega|^2.\myeqno$$ 
As far as long-time behaviours are concerned, it is sufficient to know the behaviour of
$\Lambda(\omega)$ at low frequencies (for physical reasons, $\Lambda(\omega)\to 0$ when
$\omega\to\infty$, as pictured by some cut-off function $f_c({\omega/\omega_c})$, where $\omega_c$
is linked to the bandwidth of the effectively coupled bath oscillators). In the so-called ohmic
dissipation model~[11],
$\Lambda(\omega)\sim\omega$ for $\omega\to 0$, and one sets
$$\Lambda(\omega)=m\gamma\hbar\omega f_c({\omega\over\omega_c}),\qquad\omega>0,\myeqno$$
or, equivalently,
$$K(\omega)=2\gamma f_c({\omega\over\omega_c}),\myeqno$$
where $K(\omega)=\int_{-\infty}^\infty k(t)\,e^{i\omega t}\,dt$ denotes the Fourier transform of
$k(t)$ in the ordinary sense. With a Lorentzian cut-off function
$f_c({\omega/\omega_c})={\omega_c^2/(\omega_c^2+\omega^2)}$, the retarded memory kernel
$k(t)$ is modelized by the exponential function
$$k(t)=\gamma\omega_c\,e^{-\omega_c|t|}.\myeqno$$
In Eq. (4.8), $\omega_c^{-1}$ acts as a measure of the memory time of $k(t)$.
\bigskip
{\bf 4.3. The quantum Langevin equation}

Using instead of $k(t)$ the kernel
$$\tilde k(t)=\Theta(t)\,k(t),\myeqno$$
the upper bound of the integral in Eq. (4.2) can be set equal to $+\infty$. This latter equation
can then be rewritten as
$$m\ddot x(t)+m\int_{t_i}^{+\infty}dt'\,\tilde k(t-t')\,\dot
x(t')=-m\,x(t_i)\,\tilde k(t-t_i)+F(t).\myeqno$$
Note that the real part of the Fourier transform $\tilde K(\omega)$ of $\tilde k(t)$ is linked to
$K(\omega)$ by
$$\RE\tilde K(\omega)={1\over 2}\,K(\omega).\myeqno$$

Let us assume that, at time $t=t_i$, the total density operator of the particle-plus-bath system is
factorized in the form $\rho_{\rm part.}\otimes\rho_{\rm bath}$, with $\rho_{\rm
bath}=Z^{-1}e^{-\beta H_{\rm bath}}$. The mean value of the Langevin force $F(t)$ is then equal to
zero. As for its symmetrized correlation function, 
$$C^Q_{FF}(t-t')={1\over 2}\,\left[\langle F(t)F(t')+F(t')F(t)\rangle\right],\myeqno$$
it depends only on the time difference $t-t'$ and it is given by
$$C^Q_{FF}(t-t')=m\int_{-\infty}^\infty{d\omega\over
2\pi}\,\RE\tilde K(\omega)\,\hbar\omega\,\coth{\beta\hbar\omega\over
2}\,e^{-i\omega(t-t')}.\myeqno$$
The Langevin force $F(t)$ can thus be viewed as corresponding to a stationary stochastic process.  
In the ohmic dissipation model in which $K(\omega)$ is given by Eq.~(4.7), one has, using the
Lorentzian cut-off function,
$$C^Q_{FF}(t)=m\hbar\omega_c^2\gamma\int_{-\infty}^\infty{d\omega\over
2\pi}{\omega\over\omega^2+\omega_c^2}\,\coth{\beta\hbar\omega\over
2}\,e^{-i\omega t}.\myeqno$$
The detailed analysis of the behaviour of $C^Q_{FF}(t)$ as a function of $t$ for any given 
temperature of the bath can be found in [12].

Let us now come back to the equation of motion (4.10). The r.h.s. contains, in addition to the
Langevin force $F(t)$, a term depending on the position $x(t_i)$ of the particle at time
$t_i$. The kernel $\tilde k(t)$ is supposed to have a memory time of the order of
$\omega_c^{-1}$ and we shall study the behaviour of the system for times $t$ such that
$\omega_c(t-t_i)\gg 1$. Then the quantity $\tilde k(t-t_i)$ is negligible, and the initial position
of the particle becomes irrelevant in Eq. (4.10). Mathematically, this can be realized by letting
$t_i\to -\infty$. Eq. (4.10) then takes the form of the so-called quantum Langevin equation
$$m\,{dv\over dt}+m\int_{-\infty}^{+\infty} dt'\,\tilde k(t-t')\,v(t')=F(t),\qquad v={dx\over
dt}.\myeqno$$
\vfill
\break
{\bf 4.4. The infinitely short memory limit} 

Clearly, the kernel $k(t)$ as modelized by Eq. (4.8) admits an infinitely short memory limit,
$k(t)=\gamma\delta(t)$, which is easily obtained by taking the limit $\omega_c\to\infty$. Eqs.
(4.10) or (4.15) then reduce to the non retarded equation
$$m\,{dv\over dt}+m\,\gamma\,v(t)=F(t),\myeqno$$
analogous to the classical Langevin equation (3.1), and in which memory effects are disregarded.
Note that the corresponding limit has to be taken in the Langevin force correlation function (4.14),
which then becomes independent of the modelization chosen for $k(t)$, as it should.

\mysection{The two-time velocity and displacement correlation functions of a free quantum Brownian
particle}

{\bf 5.1. The two-time velocity correlation function}

The symmetrized two-time velocity correlation function is defined by
$$C^Q_{vv}(t_1,t_2)={1\over 2}\,\left[\langle v(t_1)v(t_2)+v(t_2)v(t_1\rangle\right].\myeqno$$
As stated above, the Langevin force $F(t)$ can be viewed as corresponding to a stationary
stochastic process. Clearly, the same is true of the solution $v(t)$ of Eq. (4.15), which is valid
once the limit $t_i\to -\infty$ has been taken. Thus, Fourier analysis and the Wiener-Khintchine
theorem can be used to obtain the equilibrium velocity correlation function $C^Q_{vv}(t_1-t_2)$,
which only depends on the time difference $t_1-t_2$. Thus, as in the classical case, the velocity
of the free quantum Brownian particle is not an aging variable. 

As a result, one gets:
$$C^Q_{vv}(t)={1\over m}\int_{-\infty}^\infty{d\omega\over 2\pi}\,{\RE\tilde
K(\omega)\over|\tilde K(\omega)-i\omega|^2}\,\hbar\omega\,\coth{\beta\hbar\omega\over
2}\,e^{-i\omega t}.\myeqno$$
In the ohmic dissipation model, this expression takes the form:
$$C^Q_{vv}(t)={\gamma\over m}\int_{-\infty}^\infty{d\omega\over
2\pi}{\omega_c^2\over(\gamma\omega_c-\omega^2)^2+\omega^2\omega_c^2}\,\hbar\omega\,
\coth{\beta\hbar\omega\over 2}\,e^{-i\omega t}.\myeqno$$
In the case ${\omega_c/\gamma}>4$, which corresponds to a weak coupling between the particle and the
bath\footnote{$^2$}{The following expressions remain valid even when ${\omega_c/\gamma}<4$,
provided that the appropriate analytical continuations are done. In this case, $\omega_\pm$
has an imaginary part and therefore oscillations in time appear in the correlation
functions. We shall only here consider the weak-coupling case ${\omega_c/\gamma}>4$, since
the weak-coupling assumption is already contained in the Caldeira and Leggett Hamiltonian (4.1).},
one can write
$$C^Q_{vv}(t)={\gamma\over
m}\,\bigl(1-{4\gamma\over\omega_c}\bigr)^{-{1/2}}\int_{-\infty}^\infty{d\omega\over
2\pi}\,\left({1\over\omega^2+\omega_-^2}-{1\over\omega^2+\omega_+^2}\right)\,\hbar\omega\,
\coth{\beta\hbar\omega\over 2}\,e^{-i\omega t},\myeqno$$
where
$$\omega_\pm={\omega_c\over
2}\,\left[1\pm\bigl(1-{4\gamma\over\omega_c}\bigr)^{1/2}\right].\myeqno$$
In the infinitely short memory limit $\omega_c\to\infty$, one simply has:
$$C_{vv}^{Q,\omega_c\to\infty}(t)={\gamma\over m}\int_{-\infty}^\infty{d\omega\over
2\pi}\,{1\over\omega^2+\gamma^2}\,\hbar\omega\,\coth{\beta\hbar\omega\over
2}\,e^{-i\omega t}.\myeqno$$
At any non-zero temperature, formulas (5.4) and (5.6) can respectively be recast into the form 
$$C^Q_{vv}(t)={k_BT\over
m}\,\bigl(1-{4\gamma\over\omega_c}\bigr)^{-{1/2}}\,\left[{\gamma\over\omega_-}\,f_T(t,\omega_-)-
{\gamma\over\omega_+}\,f_T(t,\omega_+)\right]\myeqno$$
and
$$C_{vv}^{Q,\omega_c\to\infty}(t)={k_BT\over m}f_T(t,\gamma),\myeqno$$
where $f_T(t,\gamma)$ stands for the series expansion
$$f_T(t,\gamma)=e^{-\gamma|t|}+
2\,\sum_{n=1}^\infty{{e^{-\gamma|t|}-{n\over\gamma t_{\rm
th.}}e^{-{n|t|\over t_{\rm th.}}}\over 1-({n\over\gamma t_{\rm th.}})^2}}\myeqno$$
and $t_{\rm th.}$ denotes a thermal time as defined by $t_{\rm th.}={\hbar/2\pi k_BT}$.  

The detailed analysis of the behaviour of $C^Q_{vv}(t)$ as a function of $t$ can be found in [12].
Let us here just quote the following important feature: there exists a cross-over temperature $T_c$,
which is given by $T_c={\hbar\omega_-/\pi k_B}$ (or $T_c={\hbar\gamma/\pi k_B}$ in the infinitely
short memory limit), above and below which the velocity correlation function exhibits two markedly
different behaviours\footnote{$^3$}{This property can be proved by using the following
equivalent form of $f_T(t,\gamma)$:
$$f_T(t,\gamma)=\pi\gamma t_{\rm th.}\cot\pi\gamma t_{\rm
th.}\,e^{-\gamma|t|}-2\sum_{n=1}^\infty{{n\over\gamma t_{\rm
th.}}e^{-{n|t|\over t_{\rm th.}}}\over 1-({n\over\gamma t_{\rm th.}})^2}.$$}. 
Namely, for $T>T_c$, $C^Q_{vv}(t)$ is positive at
any time and therefore, despite the existence of quantum corrections, this regime may be
roughly qualified as classical. For $T<T_c$, $C^Q_{vv}(t)$ is positive at small times, passes
through a zero and is negative at large times. This regime may be qualified as quantal. 
\bigskip
{\bf 5.2. The time-dependent diffusion coefficient}

The classical definition (3.16) of the time-dependent diffusion coefficient $D(t)$ can be
readily extended to the present quantum model. Using the expression (5.3) of the symmetrized
velocity correlation function in the ohmic dissipation model, one gets for the quantum
time-dependent diffusion coefficient $D^Q(t)$ the expression
$$D^Q(t)={\gamma\over m}\int_{-\infty}^\infty{d\omega\over
2\pi}{\omega_c^2\over(\gamma\omega_c-\omega^2)^2+\omega^2\omega_c^2}\,
\hbar\,\coth{\beta\hbar\omega\over 2}\,\sin\omega t.\myeqno$$
In the short-memory limit, this expression simplifies into
$$D^{Q,\omega_c\to\infty}(t)={\gamma\over m}\int_{-\infty}^\infty{d\omega\over
2\pi}\,{1\over\omega^2+\gamma^2}\,\hbar\,\coth{\beta\hbar\omega\over 2}\,\sin\omega t.\myeqno$$
Note that formulas (5.10) and (5.11) are also valid for negative times, $D^Q(t)$ and
$D^{Q,\omega_c\to\infty}(t)$ being odd functions of $t$.

Let us now for future purpose briefly recall the main features of the behaviour of $D^Q(t)$ as a
function of $t$ for $t>0$ [12]. First, the limit value at infinite time of $D^Q(t)$ is, at any
non-zero temperature, the usual Einstein diffusion coefficient $D={k_BT/m\gamma}$. This result
holds whatever the value,  finite or not, of $\omega_c$.

For the sake of simplicity, we restrict from now on the study of the quantum time-dependent
diffusion coefficient to the short-memory limit $\omega_c\to\infty$, which presents the same gross
features as the general case, but in which the analytical calculations are more simple.  Formula
(5.11) can (at any non-zero temperature) be recast into the form of the series expansion
$$D^{Q,\omega_c\to\infty}(t)={k_BT\over m\gamma}\,\left\{1-e^{-\gamma
t}+2\,\sum_{n=1}^\infty{e^{-{nt\over t_{\rm th.}}}-e^{-\gamma t}\over 1-({n\over\gamma t_{\rm
th.}})^2}\right\},\qquad t>0.\myeqno$$ 
The quantum time-dependent diffusion coefficient thus appears as the sum of the classical
contribution
$$D^{\omega_c\to\infty}(t)={k_BT\over m}\,\bigl(1-e^{-\gamma t}\bigr),\qquad t>0,\myeqno$$
and of a supplementary contribution due to quantum effects, which can be shown to be
always positive.

Above the cross-over temperature $T_c$ as defined above, $D^{Q,\omega_c\to\infty}(t)$ increases
monotonously towards its limit value $D={k_BT/m\gamma}$ (classical regime). Below the cross-over
temperature ({\sl i.e.\/} $T<T_c$), $D^{Q,\omega_c\to\infty}(t)$ first increases, then passes
through a maximum and finally slowly decreases towards its limit value $D={k_BT/m\gamma}$ (quantum
regime). Thus, in the quantum regime, {\sl i.e.\/} when quantum effects are dominant, the
quantum time-dependent diffusion coefficient can exceed its stationary value and the diffusive
regime is only attained very slowly, namely after times
$t\gg t_{\rm th.}$. At $T=0$, the quantum time-dependent diffusion coefficient can be expressed in
terms of exponential integral functions [14] as
$$D^{Q,\omega_c\to\infty}_{T=0}(t)={\hbar\over 2\pi m\gamma}\,\left[e^{-\gamma t}\,\overline{\rm
Ei}(\gamma t)-e^{\gamma t}\,{\rm Ei}(-\gamma t)\right],\qquad t>0.\myeqno$$
$D^{\omega_c\to\infty}_{T=0}(t)$ passes through a maximum at a time
$t_m(T=0)\sim\gamma^{-1}$. For $\gamma t\gg 1$, one has:
$$ D^{Q,\omega_c\to\infty}_{T=0}(t)\sim{\hbar\over\pi m}{1\over\gamma t},\qquad\gamma t\gg
1.\myeqno$$

The behaviour of the quantum time-dependent diffusion coefficient\break $D^{Q,\omega_c\to\infty}(t)$
for several different temperatures on both sides of $T_c$ is illustrated on Fig.~2. Interestingly
enough, these curves never intersect. One can indeed easily check that the slopes at the origin
${dD^{Q,\omega_c\to\infty}(t)/dt}|_{t=0}=C^{Q,\omega_c\to\infty}_{vv}(t=0)$\break increase with $T$,
as well as the asymptotic values at infinite time $D={k_BT/m\gamma}$. One can also show that, at
any fixed time $t$, the quantum time-dependent diffusion coefficient is a monotonously increasing
function of the temperature\footnote{$^4$}{See Section 6 (Eq. (6.12)).} (Fig.~3). This physically
expected property, which is valid in the whole temperature range and thus in the quantum regime
({\sl i.e.\/} $T>T_c$) as well as in the classical one ({\sl i.e.\/}
$T<T_c$), will play an important role in the following.    
\bigskip
{\bf 5.3. The two-time displacement correlation function and the response function}

Let us consider the symmetrized two-time
displacement correlation function $C^Q_{xx}(t,t';t_0)$, as defined by 
$$C^Q_{xx}(t,t';t_0)={1\over
2}\,\left\{\langle[x(t)-x(t_0)][x(t')-x(t_0)]\rangle+\langle[x(t')-x(t_0)][x(t)-x(t_0)]\rangle
\right\},\myeqno$$
where $x(t)$ is the free quantum Brownian particle position operator.
Still assuming that the limit $t_i\to -\infty$ has been taken from the beginning,
$C^Q_{xx}(t,t';t_0)$ can be obtained from the equilibrium symmetrized velocity correlation function
$C^Q_{vv}(t)$ by computing the double integral
$$C^Q_{xx}(t,t';t_0)=\int_{t_0}^t dt_1\int_{t_0}^{t'}dt_2\,C^Q_{vv}(t_1-t_2).\myeqno$$
However, the detailed expression of $C^Q_{xx}(t,t';t_0)$ is of no use in the following analysis,
which will be seen in the next section to rely only  on the expression of the quantum time-dependent
diffusion coefficient $D^Q(t)$, linked to $C^Q_{xx}(t,t;0)$ by the relation
$$D^Q(t)={1\over 2}\,{d\over dt}\,C^Q_{xx}(t,t;0).\myeqno$$
which is the quantum analog of the classical equation (3.15).

As for the displacement response function $\chi_{xx}(t,t')$, it can be obtained from the equation
of motion (4.15) in which one adds to the Langevin force a non random force proportional to
$\delta(t-t')$. One gets, in the case\footnote{$^5$} {Formula (5.19) with
$\omega_\pm$ as given by Eq. (5.5) remains valid even when\break
${\omega_c/\gamma}<4$, provided that the appropriate analytical continuations are done.}
${\omega_c/\gamma>4}$, and for any bath temperature,
$$\displaylines{\chi_{xx}(t,t')=\Theta(t-t')\,{1\over
m\gamma}\times\hfill\cr
\hfill\left[1-\bigl(1-{4\gamma\over\omega_c}\bigr)^{-{1/2}}\,{\omega_+^2\over\omega_c^2}\,
e^{-\omega_-(t-t')}+
\bigl(1-{4\gamma\over\omega_c}\bigr)^{-{1/2}}\,{\omega_-^2\over\omega_c^2}\,
e^{-\omega_+(t-t')}\right].\mydispeqno}$$ For
well-separated time scales $\omega_c^{-1}$ and $\gamma^{-1}$ ($\omega_c^{-1}\ll\gamma^{-1}$), one
has simply:
$$\chi_{xx}(t,t')=\Theta(t-t')\,{1\over
m\gamma}\,\bigl[1-e^{-\gamma(t-t')}+{\gamma^2\over\omega_c^2}\,e^{-\omega_c(t-t')}\bigr].\myeqno$$
In the infinitely short memory limit $\omega_c\to\infty$, one recovers as expected for
$\chi_{xx}(t,t')$ the expression (3.10) of the Langevin model. 

\mysection{Modified quantum fluctuation-dissipation theorem in the presence of aging}

In the presence of aging effects, that is, when the dynamic variable of interest does not
equilibrate with the bath, even at large times, the quantum fluctuation-dissipation theorem
(QFDT) cannot be applied in its standard form. A modified form of QFDT in the non-equilibrium case
allowing for the description of aging effects has been proposed in [7] for mean-field spin-glass
models. Here, we propose a modified form of QFDT which can conveniently be applied to the case of
the displacement of the free quantum Brownian particle, {\sl i.e.\/} to the problem of diffusion.
\bigskip
{\bf 6.1. Relation between the displacement response function and the time-dependent diffusion
coefficient}

First, we shall refer to a time-dependent formulation of the equilibrium QFDT, namely the relation
$$\chi_{BA}(t,t')={2\over\pi\hbar}\,\Theta(t-t')\,\int_{-\infty}^\infty dt''\,{\partial
C_{BA}(t,t'')\over\partial t''}\,\left[-\log\Bigl|\tanh{\pi(t''-t')\over
2\beta\hbar}\Bigr|\right],\myeqno$$
and its reciprocal,
$${\partial C_{BA}(t,t')\over\partial t'}=-{\hbar\over 2\pi}\,{\rm vp}\int_{-\infty}^\infty
dt''\,{\partial\over\partial t''}\bigl[\chi_{BA}(t,t'')-\chi_{AB}(t'',t)\bigr]\,{\pi\over\beta\hbar}
\,\coth{\pi(t''-t')\over\beta\hbar},\myeqno$$
which are proved in the Appendix\footnote{$^6$}{For the sake of simplicity, the superior
indexes $Q$ in the correlation functions have been dropped in the general formulas.}.

Then, let us turn to the specific problem of the diffusion of a free quantum Brownian
particle. Since the limit $t_i\to -\infty$ has been taken from the beginning, the particle velocity
has reached equilibrium. The equilibrium QFDT (6.1) can thus be applied to this dynamic variable,
yielding, with obvious notations,
$$\chi_{vx}(t_1,t')={2\over\pi\hbar}\,\Theta(t_1-t')\,\int_{-\infty}^\infty dt''\,{\partial
C^Q_{vx}(t_1-t'')\over\partial t''}\,\left[-\log\Bigl|\tanh{\pi(t''-t')\over
2\beta\hbar}\Bigr|\right],\myeqno$$
that is
$$\chi_{vx}(t_1,t')={2\over\pi\hbar}\,\Theta(t_1-t')\,\int_{-\infty}^\infty dt''\,
C^Q_{vv}(t_1-t'')\,\left[-\log\Bigl|\tanh{\pi(t''-t')\over
2\beta\hbar}\Bigr|\right].\myeqno$$
Since the displacement response function $\chi_{xx}(t,t')$ can be expressed as the integral
$$\chi_{xx}(t,t')=\int_{t'}^t\chi_{vx}(t_1,t')\,dt_1,\myeqno$$
one gets for this quantity, making use of the identity (deduced from Eq. (5.17))
$${\partial C^Q_{xx}(t,t'';t_0)\over\partial
t''}\Bigr|_{t_0=t'}=\int_{t'-t''}^{t-t''}du\,C^Q_{vv}(u),\myeqno$$
the expression
$$\chi_{xx}(t,t')={2\over\pi\hbar}\,\Theta(t-t')\,\int_{-\infty}^\infty dt''\,{\partial
C^Q_{xx}(t,t'';t_0)\over\partial t''}\Bigr|_{t_0=t'}\,\left[-\log\Bigl|\tanh{\pi(t''-t')\over
2\beta\hbar}\Bigr|\right].\myeqno$$
Note that writing formula (6.7) amounts to make use of the same argument as in the classical
case, namely to apply QFDT as expressed by Eq. (6.1) to $x(t)-x(t')$. 

Since one has, in terms of the quantum time-dependent diffusion coefficient as defined by Eq.
(5.18),
$${\partial
C^Q_{xx}(t,t'';t_0)\over\partial t''}\Bigr|_{t_0=t'}=D^Q(t-t'')+D^Q(t''-t'),\myeqno$$
one can rewrite Eq. (6.7) as
$$\chi_{xx}(t,t')={2\over\pi\hbar}\,\Theta(t-t')\,\int_{-\infty}^\infty
dt''\,[D^Q(t-t'')+D^Q(t''-t')]\,\left[-\log\Bigl|\tanh{\pi(t''-t')\over
2\beta\hbar}\Bigr|\right],\myeqno$$
that is, $D^Q(t)$ being an odd function of $t$,
$$\chi_{xx}(t,t')={2\over\pi\hbar}\,\Theta(t-t')\,\int_{-\infty}^\infty dt''
D^Q(t-t'')\,
\left[-\log\Bigl|\tanh{\pi(t''-t')\over 2\beta\hbar}\Bigr|\right].\myeqno$$
One can also write equivalently 
$$\chi_{xx}(t,t')={2\over\pi\hbar}\,\Theta(t-t')\,\int_{-\infty}^\infty dt_1\,D^Q(t-t'-t_1)\,
\left[-\log\Bigl|\tanh{\pi t_1\over 2\beta\hbar}\Bigr|\right],\myeqno$$
a formulation which clearly displays the convolution product structure of the displacement response
function, when expressed in terms of the quantum time-dependent diffusion coefficient. 

Inverting this convolution product amounts to write the reciprocal rela-\break tion (6.2) for the
specific problem of the diffusion of a free quantum Brownian particle, that is:
$$D^Q(t-t')=-{\hbar\over 2\pi}\,{\rm vp}\int_{-\infty}^\infty
dt''{\partial\over\partial
t''}[\chi_{xx}(t,t'')-\chi_{xx}(t'',t)]\,{\pi\over\beta\hbar}\,
\coth{\pi(t''-t')\over\beta\hbar}.\myeqno$$
Since the displacement response function $\chi_{xx}$ does not depend on the bath temperature (Eqs.
(5.19) or (5.20)), Eq. (6.12) displays the above quoted property that, at any fixed time $t$,
$D^Q(t)$ is a monotonously increasing function of the temperature. In the infinitely short memory
limit, taking into account the corresponding expression (3.10) of 
$\chi_{xx}$, one gets from Eq. (6.12):
$$\displaylines{D^{Q,\omega_c\to\infty}(t-t')={\hbar\over 2\pi m}\,{\rm
vp}\int_{-\infty}^\infty
dt_1\hfill\cr
\hfill[\Theta(t-t'-t_1)e^{-\gamma(t-t'-t_1)}+\Theta(t_1-t+t')e^{\gamma(t-t'-t_1)}]\,
{\pi\over\beta\hbar}\,\coth{\pi t_1\over\beta\hbar}.\mydispeqno\cr}$$

Eqs. (6.11) or (6.12) are the quantum generalization of the classical equation (3.22)
and reduce to it in the classical limit, which can be easily checked by making use of the property
$$-\log|\tanh\,ax|\sim_{a\to\infty}{\pi^2\over 4a}\,\delta(x).\myeqno$$
Exactly like its classical analog Eq. (3.22), Eq. (6.11) allows to express the displacement response
function $\chi_{xx}(t,t')$ in terms of the time-dependent diffusion coefficient.
However, contrary to the classical case in which $\chi_{xx}(t,t')$ is directly proportional to
$D(t-t')$, in the quantum formulation $\chi_{xx}(t,t')$ is a convolution product, for the value
$t-t'$ of the argument, of the functions $D^Q(t_1)$ and $-\log|\tanh({\pi t_1/2\beta\hbar})|$. This
latter function is very peaked around $t_1=0$ at high temperature while it becomes more and more
spread around this value as the temperature decreases.
\bigskip
{\bf 6.2. The modified quantum fluctuation-dissipation theorem}

Proceeding in the same way as in the classical case, let us write a modified QFDT relating the
displacement response function $\chi_{xx}$ and the derivative ${\partial
C^Q_{xx}(t,t';t_0)/\partial t'}$ of the correlation function
$C^Q_{xx}$ as defined by Eq. (5.16), with $t_0\leq t'<t$, as
$$\displaylines{{\partial C^Q_{xx}(t,t';t_0)\over\partial t'}=\hfill
\cr\hfill  -{\hbar\over 2\pi}\,{\rm
vp}\int_{-\infty}^\infty dt''{\partial\over\partial
t''}[\chi_{xx}(t,t'')-\chi_{xx}(t'',t)]\,{\pi\over\beta X^Q\hbar}\,
\coth{\pi(t''-t')\over\beta X^Q\hbar},\mydispeqno\cr}$$ 
where, for short, $X^Q$ stands for
$X^Q(t,t';t_0)$. Using the quantum analog of the classical equation (3.24), one gets from
Eq. (6.15):
$$\displaylines{D^Q(t-t')+D^Q(t'-t_0)=\hfill\cr
\hfill -{\hbar\over 2\pi}\,{\rm vp}\int_{-\infty}^\infty
dt''{\partial\over\partial
t''}[\chi_{xx}(t,t'')-\chi_{xx}(t'',t)]\,{\pi\over\beta X^Q\hbar}\,
\coth{\pi(t''-t')\over\beta X^Q\hbar}.\mydispeqno\cr}$$ 
The quantity $X^Q(t,t';t_0)$ in Eqs. (6.15)-(6.16) represents the QFDT violation
factor. 

The parameter $\beta^Q_{\rm eff.}(t,t';t_0)$ linked to $X^Q(t,t';t_0)$ by the equation 
$$\beta^Q_{\rm eff.}(t,t';t_0)=\beta\,X^Q(t,t';t_0),\myeqno$$
which is the quantum analog of the classical equation (2.10), can be interpreted as an associated
effective inverse temperature, defined for $\tau=t-t'>0$ and $t_w=t'-t_0>0$. In terms of
this quantity, Eq. (6.16) can be rewritten as
$$\displaylines{D^Q(t-t')+D^Q(t'-t_0)=\hfill\cr
\hfill -{\hbar\over 2\pi}\,{\rm vp}\int_{-\infty}^\infty
dt''{\partial\over\partial
t''}[\chi_{xx}(t,t'')-\chi_{xx}(t'',t)]\,{\pi\over\beta^Q_{\rm eff.}\hbar}\,
\coth{\pi(t''-t')\over\beta^Q_{\rm eff.}\hbar},\mydispeqno\cr}$$
where for short $\beta^Q_{\rm eff.}$ stands for $\beta^Q_{\rm eff.}(t,t';t_0)$. 

Let us add as a comment that, like its classical analog, the parameter\break $T_{\rm
eff.}^Q={1/\beta^Q_{\rm eff.}}$, if it allows to write the modified QFDT (6.18), cannot however be
considered as a {\sl bona fide\/} physical temperature. In particular, it does not control the
thermalization of the free quantum Brownian particle (which is effective here since the limit
$t_i\to -\infty$ has been taken from the very beginning). 
\bigskip
{\bf 6.3. Determination of the effective temperature}

When compared with the r.h.s. of Eq. (6.12), the r.h.s. of Eq. (6.18) appears to be formally 
identical to the quantum time-dependent diffusion coefficient
$D^Q_{T^Q_{\rm eff.}}(t-t')$ at the effective temperature
$T^Q_{\rm eff.}={1/k_B\beta^Q_{\rm eff.}}$. One can thus rewrite Eq. (6.18) as
$$D^Q(t-t')+D^Q(t'-t_0)=D^Q_{T^Q_{\rm eff.}}(t-t').\myeqno$$
Eq. (6.19) in turn allows for the determination of the effective temperature $T^Q_{\rm
eff.}$ as a function of $t-t'$ and $t'-t_0$. Since the quantum time-dependent diffusion coefficient
is, at any time, a monotonously increasing function of the temperature, Eq. (6.19) yields for
$T^Q_{\rm eff.}(t-t',t'-t_0)$ a uniquely defined value.

In the classical limit in which the time-dependent diffusion coefficient is proportional to the
temperature, it is easy to check that the effective inverse temperature deduced in this limit from
Eq. (6.19) verifies the equation
$$\beta_{\rm eff.}(t,t';t_0)=\beta\,{D(t-t')\over D(t-t')+D(t'-t_0)},\myeqno$$
in accordance with Eq. (3.19) with $\tau=t-t'$ and $t_w=t'-t_0$.

Note that, except in the classical limit, $\beta^Q_{\rm eff.}$ as
defined by Eq. (6.19) is not simply proportional to $\beta$, so that the QFDT violation cannot be
described by a simple rescaling of the bath temperature.
\bigskip
{\bf 6.4. Analysis of the behaviour of the effective temperature in the ohmic
dissipation model}

In contrast with their classical counterparts, $X^Q(t,t';t_0)$ and $\beta^Q_{\rm eff.}(t,t';t_0)$
are not given by a simple ratio of time-dependent diffusion coefficients like Eq. (3.18) or (3.19),
but have to be deduced from the implicit equation (6.19), which in general can only be solved
numerically. Let us restrict the study, for the sake of simplicity, to the short-memory limit
$\omega_c\to\infty$. Solving Eq. (6.19) amounts then to find the solution $\beta^Q_{\rm
eff.}(t,t';t_0)$ of the equation
$$\displaylines{\int_{-\infty}^\infty{d\omega\over
2\pi}\,{1\over\omega^2+\gamma^2}\,\hbar\,\coth{\beta^Q_{\rm eff.}(t,t';t_0)\hbar\omega\over
2}\,\sin\omega(t-t')=\hfill\cr
\hfill\int_{-\infty}^\infty{d\omega\over
2\pi}\,{1\over\omega^2+\gamma^2}\,\hbar\,\coth{\beta\hbar\omega\over
2}\,[\sin\omega(t-t')+\sin\omega(t'-t_0)].\mydispeqno\cr}$$
This solution is a function of the time difference $\tau=t-t'$ and of the waiting time
$t_w=t'-t_0$.

The curves representing $X^Q(\tau,t_w)={\beta^Q_{\rm eff.}(\tau,t_w)/\beta}$ as a function of 
$\tau$ for various non zero values of $t_w$ are plotted on Figs.~5 to 7 for several non
zero bath temperatures on both sides of $T_c$. Note that, interestingly enough, Eqs. (6.19) and
(6.21) also allow to define the effective temperature when $T=0$. The corresponding curves
are plotted on Fig.~8. 

Let us now comment upon the results obtained (Figs.~5 to 8).  For all values of $\tau$,
and at any non zero temperature, $X^Q(\tau,t_w)$ is equal to 1 when $t_w=0$ and it differs from~1
when $t_w\neq 0$. Thus, like in the classical case, aging is always present as far as the
displacement of the free quantum Brownian particle is concerned. For a given value of
$t_w$, $X^Q(\tau,t_w)$ and $\beta^Q_{\rm eff.}(\tau,t_w)$ increase monotonously with $\tau$ towards
a finite limit $X^Q_\infty(t_w)$ (Figs. 5 to 7). At $T=0$, $T^Q_{\rm eff.}(\tau,t_w)$
decreases monotonously with $\tau$ towards zero (Fig. 8).

Let us discuss the limit value of $X^Q$ for large values of $\tau$ ($\tau\gg\gamma^{-1}, t_{\rm
th.}$). For a given $t_w$, one has
$$X^Q_\infty(t_w)\equiv\lim_{\tau\gg\gamma^{-1},t_{\rm th.}}X^Q(\tau,t_w)={1\over 1+{m\gamma\over
k_BT}\,D^Q(t_w)}.\myeqno$$
In the classical regime ({\sl i.e.\/} $T>T_c$), $D^Q(t_w)$ is a monotonously increasing function of
$t_w$, which is reflected in the fact that $X^Q_\infty(t_w)$ decreases monotonously with $t_w$ (see
Fig.~5 for $T=2T_c$). In the quantum regime ({\sl i.e.\/} $T<T_c$), $D^Q(t_w)$ is not a monotonous
function of $t_w$. Consequently, the same is true of $X^Q_\infty(t_w)$ (see Fig.~7 for
$T={T_c/2}$). A similar phenenomenon takes place at $T=0$, in which case, for any given value of
$\tau$,  $T^Q_{\rm eff.}(\tau,t_w)$ first increases with $t_w$, passes through a maximum for a
value $t_{wm}\sim\gamma^{-1}$, then decreases towards zero (Fig.~8).
\bigskip
{\bf 6.5. Discussion}

The parameter $\beta^Q_{\rm eff.}(t,t';t_0)$ once determined as described above,
one can write, using Eqs. (6.18) and (6.19),
$$D^Q_{T^Q_{\rm eff.}}(t-t')= -{\hbar\over 2\pi}\,{\rm vp}\int_{-\infty}^\infty
dt''{\partial\over\partial
t''}[\chi_{xx}(t,t'')-\chi_{xx}(t'',t)]\,{\pi\over\beta^Q_{\rm eff.}\hbar}\,
\coth{\pi(t''-t')\over\beta^Q_{\rm eff.}\hbar}.\myeqno$$

On another hand, coming back to the expression (6.10) of $\chi_{xx}(t,t')$, and taking advantage of
the fact that the displacement response function does not actually depend on $\beta$ (Eqs. (5.19)
or (5.20)), one may replace $\beta$ by $\beta^Q_{\rm eff.}(t-t',t'-t_0)$ in this expression,
which yields 
$$\chi_{xx}(t,t')={2\over\pi\hbar}\,\Theta(t-t')\,\int_{-\infty}^\infty dt''
D^Q_{T^Q_{\rm eff.}}(t-t'')\,
\left[-\log\Bigl|\tanh{\pi(t''-t')\over 2\beta^Q_{\rm eff.}\hbar}\Bigr|\right].\myeqno$$
Eqs. (6.23) ({\sl resp.\/} (6.24)) are formally similar to Eqs. (6.10) ({\sl resp.\/} (6.12)),
except for the fact that, due to waiting time effects, the diffusing particle is considered as
being in contact with a bath at $T^Q_{\rm eff.}$, an effective temperature depending on both
time arguments $t-t'$ and $t'-t_0$. 

Let us emphasize however that, in contrast with the sets of equations ((6.1), (6.2)) or
((6.10), (6.12)) which are reciprocal convolution relations, Eq. (6.24) cannot be viewed as
the reciprocal relation of Eq. (6.23), since these two equations do not involve the same functions.
Actually, the convolution product structure is lost in Eqs. (6.23) and (6.24), because of the
dependence of
$T^Q_{\rm eff.}$ and $\beta^Q_{\rm eff.}$ on $t-t'$ and $t'-t_0$. As a result, while Eq. (6.23)
links the displacement response function and $D^Q_{T^Q_{\rm eff.}(t-t',t'-t_0)}(t-t')$, Eq. (6.24)
links the displacement response function with a different function $D^Q_{T^Q_{\rm
eff.}(t-t',t'-t_0)}(t-t'')$. The dependence on the running time variable $t'$ in $D^Q_{T^Q_{\rm
eff.}(t-t',t'-t_0)}(t-t')$ is actually different from the
$t''$ dependence of $D^Q_{T^Q_{\rm eff.}(t-t',t'-t_0)}(t-t'')$.
In particular, except for $t''=t'$, one has 
$$D^Q(t-t'')+D^Q(t''-t_0)\neq D^Q_{T^Q_{\rm eff.}}(t-t'').\myeqno$$

\mysection{Quantum aging and diffusion}

Keeping in mind the idea according to which aging effects in diffusion are basically a
consequence of the fact that fluctuations in the particle displacement do take place even during
the waiting time, a description of these effects in quantum
diffusion can be obtained by studying the ratio
$$\tilde X^Q(\tau,t_w)={D^Q(\tau)\over D^Q(\tau)+D^Q(t_w)},\qquad\tau>0,\qquad t_w\geq 0.\myeqno$$
In the classical limit, this quantity is nothing but the FDT violation factor $X(\tau,t_w)$ (Eq.
(3.18)). However, outside this limit, $\tilde X^Q(\tau,t_w)$ is not equal to the QFDT violation
factor
$X^Q(\tau,t_w)={\beta^Q_{\rm eff.}(\tau,t_w)/\beta}$, as defined by Eq. (6.19).

Note that, in terms of $\tilde X^Q(t-t',t'-t_0)$, Eq. (6.12) can be rewritten under
the form of a relation between the displacement response function $\chi_{xx}$ and the derivative
${\partial C^Q_{xx}(t,t';t_0)/\partial t'}$($t_0\leq t'< t$), namely
$$\displaylines{\tilde X^Q(t,t';t_0)\,{\partial C^Q_{xx}(t,t';t_0)\over\partial
t'}=\hfill\cr
\hfill -{\hbar\over 2\pi}\,{\rm vp}\int_{-\infty}^\infty
dt''{\partial\over\partial
t''}[\chi_{xx}(t,t'')-\chi_{xx}(t'',t)]\,{\pi\over\beta\hbar}\,
\coth{\pi(t''-t')\over\beta\hbar}.\mydispeqno\cr}$$
This equation, which generalizes the classical formula (2.9), can be viewed as an alternative way
of writing a modified QFDT.

The curves representing $\tilde X^Q(\tau,t_w)$ as a function of $\tau$ for various non zero values
of $t_w$ are plotted on Figs. 9 to 11 for $T=T_c$, $T={T_c/2}$ and $T=0$. 

Let us now comment upon the results obtained (Figs.~9 to 11). For a given value of $t_w$, the
analysis of $\tilde X^Q(\tau,t_w)$ closely follows the one for $D^Q(\tau)$. Thus, in the classical
regime ({\sl i.e.\/} $T>T_c$), $\tilde X^Q(\tau,t_w)$ increases monotonously towards its limit
value. This is also the case when $T=T_c$ (Fig. 9). In the quantum regime ({\sl i.e.\/} $T<T_c$), it
passes through a maximum before attaining its limit (Figs. 10 and 11). 

The limit value of $\tilde X^Q$ for large values of $\tau$  ($\tau\gg\gamma^{-1}, t_{\rm
th.}$) is
$$\tilde X^Q_\infty(t_w)\equiv\lim_{\tau\gg\gamma^{-1},t_{\rm th.}}\tilde X^Q(\tau,t_w)={1\over
1+{m\gamma\over k_BT}\,D^Q(t_w)},\myeqno$$
Thus, one has:
$$\tilde X^Q_\infty(t_w)=X^Q_\infty(t_w).\myeqno$$
As indicated above, in the quantum regime, this quantity is not a monotonous function of $t_w$ 
(see Fig.~10 for $T={T_c/2}$). The limit $\tilde X^Q_\infty(t_w)$ is attained after a time
$\tau\gg\gamma^{-1},t_{\rm th.}$ ($t_{\rm th.}={\hbar/2\pi k_BT}$), that is, for $T>T_c$,
$\tau\gg\gamma^{-1}$, and, for $T<T_c$, $\tau\gg t_{\rm th.}$. This latter time becomes
larger and larger as the temperature decreases. 

At $T=0$, $t_{\rm th.}$ diverges,
and a singular behaviour takes place. The diffusion is anomalous, as pictured by the fact that
$D^Q(t)$ decreases as $t^{-1}$ at large time ($\gamma t\gg 1$). For large values of $\tau$ and
$t_w$ ($\tau\gg\gamma^{-1}$, $t_w\gg\gamma^{-1}$), one has (Fig.~11):
$$\tilde X^Q(\gamma\tau\gg 1,\gamma t_w\gg 1)\simeq{t_w\over\tau+t_w}.\myeqno$$
Note that, in contradistinction with what happens at finite temperature, the zero-temperature
quantity $\tilde X^Q$ possesses a singular character : as displayed by Eq.~(7.5), the two limits
$\tau\to\infty$ and $t_w\to\infty$ do not commute. Particularly noteworthy is the fact that,
although in the limit $t_w\to\infty$, $\tilde X^Q$ tends towards 1
(Eq. (7.5)), the Brownian particle displacement $x(t)-x(t_0)$ is
not at thermal equilibrium, even though diffusion is slowed down:
$$\langle[x(t)-x(t_0)]^2\rangle\simeq 2\,{\hbar\over
\pi m\gamma}\,\log\gamma(t-t_0),\qquad\gamma(t-t_0)\gg 1.\myeqno$$

\mysection{Conclusion}

This paper was devoted to the study of some aspects of a very fundamendal problem of non-equilibrium
quantum statistical mechanics, namely the quantum Brownian motion of a free particle, as described
in the framework of the Caldeira and Leggett ohmic dissipation model. We were concerned by
the behaviour of two-time correlation functions, especially the Brownian particle velocity and
displacement correlation functions $C^Q_{vv}$ and $C^Q_{xx}$. Let us just here sum up some
important features of the results.

At a certain time $t_i$, the particle is set in contact with a thermal bath at temperature $T$. In
the limit $t_i\to -\infty$, at any finite time $t$ the particle has been in contact with
the bath during an infinite interval of time, and its velocity $v(t)$ is in thermal equilibrium.
Correspondingly, the velocity correlation function $C^Q_{vv}(t_1,t_2)$ only depends on the
difference $t_1-t_2$ of the two times involved. A quantum time-dependent diffusion coefficient can
then be defined as $D^Q(t)=\int_0^t C^Q_{vv}(u)du$. For any given value of $t$, $D^Q(t)$
is a monotonously increasing function of the bath temperature, as expected on physical grounds. 

Still assuming that the limit $t_i\to -\infty$ has been taken from the very beginning, the situation
for the particle displacement $x(t)-x(t_0)$ ($t_0\leq t$) is completely different. Actually, since
diffusion is going on, the variable $x(t)-x(t_0)$ never attains equilibrium. Consequently, for any
times $t$ and $t'$ such that $t_0\leq t\leq t'$, the displacement correlation function
$C^Q_{xx}(t,t';t_0)$ depends both on the time difference $\tau=t-t'$ and on the waiting time or age
$t_w=t'-t_0$. The above analysis holds whatever the temperature of the bath.  

Thus, in the case of diffusion, the dynamic variable of interest, namely the particle displacement,
does not equilibrate with the bath, even at large times, so that the equilibrium QFDT does not
hold. In order to discuss at best how a modified quantum fluctuation-dissipation theorem (QFDT) can
be written for the diffusion problem, we have chosen to refer to a time-dependent formulation of the
equilibrium QFDT, which, in general notations, links {\sl via\/} convolution relations a response
function
$\chi_{BA}(t,t')$ and the partial derivative ${\partial C_{BA}(t,t')/\partial t'}$ of the associated
correlation function. We have shown that it is indeed possible to write a modified
QFDT relating the displacement response function $\chi_{xx}$ to the partial derivative
${\partial C^Q_{xx}(t,t';t_0)/\partial t'}=D^Q(\tau)+D^Q(t_w)$, provided that one introduces an
associated effective inverse temperature $\beta^Q_{\rm eff.}$ depending on both time arguments
$\tau$ and $t_w$. Except in the classical limit,  $\beta^Q_{\rm eff.}$ is not simply proportional to
$\beta={1/k_BT}$, so that the QFDT violation cannot be reduced to a simple rescaling of the bath
temperature.  

The definition of $\beta^Q_{\rm eff.}$ heavily relies on the quantum time-dependent
diffusion coefficient $D^Q$. More specifically, $\beta^Q_{\rm eff.}(\tau,t_w)$, which
roughly speaking takes into account even those fluctuations in the particle displacement which take
place during the waiting time, can be computed from the quantum time-dependent diffusion
coefficients $D^Q(\tau)$ and $D^Q(t_w)$. Interestingly enough, the fact that $\beta^Q_{\rm eff.}$
is a uniquely defined function of $\tau$ and $t_w$ finds its origin in the fact that, at any given
time, $D^Q_{\rm eff.}$ is a monotonously increasing function of the bath temperature. The analysis
of $\beta^Q_{\rm eff.}$ shows that, at any bath temperature, and even at $T=0$, aging is always
present as far as the displacement of the free quantum Brownian particle is concerned. For any
given value of $t_w$, the effective inverse temperature $\beta^Q_{\rm eff.}(\tau,t_w)$ is a
monotonously increasing function of $\tau$. However, the parameter $T^Q_{\rm eff.}={1/\beta^Q_{\rm
eff.}}$, albeit it allows to write a modified QFDT for the variable $x(t)-x(t_0)$, cannot be given
the full significance of a physical temperature, since in particular it does not control the
thermalization of the free Brownian particle (a phenomenon which is effective for a particle which
has been set in contact with the thermal bath for an infinite interval of time).

As a natural generalization of this study, we intend, still remaining in the quantum framework, to
consider other models for the dissipation, in which anomalous diffusive behaviours are known to
take place [16], and to study the related two-time correlation functions and aging properties.

\vfill
\break
\noindent
{\smalltitle Appendix: time-dependent formulation of the equilibrium QFDT} 
\bigskip
The equilibrium quantum fluctuation-dissipation theorem (QFDT) is usually written in a form which
involves generalized susceptibilities and spectral densities, which are frequency-dependent
quantities. In order to characterize at best the modifications of QFDT due to aging effects, it is
clearly more convenient to have at hand a formulation of the theorem in the time domain. 
Such a formulation establishes the link between the response function $\chi_{BA}(t,t')$ and the
symmetrized correlation function $C_{BA}(t,t')={1\over 2}[\langle A(t')B(t)\rangle+\langle
B(t)A(t')\rangle]$ or its derivative ${\partial C_{BA}(t,t')/\partial t'}$. 

In order to establish this link, let us consider two quantum-mechanical observables $A$ and $B$ with
thermal equilibrium correlation functions verifying the detailed balance property
$$\langle A(t'-i\hbar\beta)B(t)\rangle=\langle B(t)A(t')\rangle,\eqno(A.1)$$
and let us compute the contour integral 
$$I=\oint_{\Gamma}\langle
A(z)B(t)\rangle\,{\pi\over\beta\hbar}\,{1\over\sinh{\pi(z-t')\over\beta\hbar}}\,dz\eqno(A.2)$$
where $\Gamma$ is the closed contour represented on Fig.~4, with no singularities inside. This
integral is equal to zero, according to the Cauchy theorem. In the limit $R\to\infty$, one obtains,
by gathering the various contributions to $I$, the relation
$$\displaylines{i\pi[\langle B(t)A(t')\rangle-\langle A(t')B(t)\rangle]=\hfill\cr
\hfill{\rm
vp}\int_{-\infty}^\infty dt''[\langle
A(t'')B(t)\rangle+\langle
B(t)A(t'')\rangle]\,{\pi\over\beta\hbar}\,{1\over\sinh{\pi(t''-t')\over\beta\hbar}},
\qquad(A.3)\cr}$$
where the symbol ${\rm vp}$ denotes the Cauchy principal value. Taking into account the Kubo formula
for the response function,
one gets from Eq. ($A.3$):
$$\chi_{BA}(t,t')={2\over\pi\hbar}\,\Theta(t-t')\,{\rm vp}\int_{-\infty}^\infty dt''
C_{BA}(t,t'')\,{\pi\over\beta\hbar}\,{1\over\sinh{\pi(t''-t')\over\beta\hbar}}.\eqno(A.4)$$
Eq. ($A$.4) represents a formulation of QFDT in the time domain. It allows to compute the response
function $\chi_{BA}$ in terms of the symmetrized correlation function $C_{BA}$. 
Integrating by parts, one can recast this formula into the equivalent form
$$\chi_{BA}(t,t')={2\over\pi\hbar}\,\Theta(t-t')\,\int_{-\infty}^\infty dt''\,{\partial
C_{BA}(t,t'')\over\partial t''}\,\left[-\log\Bigl|\tanh{\pi(t''-t')\over
2\beta\hbar}\Bigr|\right].\eqno(A.5)$$

In the particular case $A=B$, Eq. ($A$.5) represents the quantum analog of the
classical formula (2.8). It reduces to this latter formula in the classical limit.

Conversely, computing on the same contour (Fig.~4) the integral
$$J=\oint_{\Gamma}\langle A(z)B(t)\rangle\,{\pi\over\beta\hbar}\,\coth{\pi(z-t')\over\beta\hbar}\,dz,\eqno(A.6)$$
one gets
$$\displaylines{{1\over 2}[\langle A(t')B(t)\rangle+\langle B(t)A(t')\rangle]=\hfill\cr
\hfill -{i\over 2\pi}\,{\rm
vp}\int_{-\infty}^\infty dt''\,[\langle
B(t)A(t'')\rangle-\langle A(t'')B(t)\rangle]\,{\pi\over\beta\hbar}\,\coth{\pi(t''-t')\over\beta\hbar},\qquad(A.7)\cr}$$
that is
$$C_{BA}(t,t')=-{\hbar\over 2\pi}\,{\rm vp}\int_{-\infty}^\infty
dt''\,\bigl[\chi_{BA}(t,t'')-\chi_{AB}(t'',t)\bigr]\,{\pi\over\beta\hbar}
\,\coth{\pi(t''-t')\over\beta\hbar}.\eqno(A.8)$$
Eq. ($A.8$) represents another formulation of QFDT in the time domain. It can be viewed as the
reciprocal of Eq. ($A.4$), since it allows to compute the symmetrized correlation function $C_{BA}$
in terms of the dissipative part of the response function~$\chi_{BA}$. One has also:
$${\partial C_{BA}(t,t')\over\partial t'}=-{\hbar\over 2\pi}\,{\rm vp}\int_{-\infty}^\infty
dt''\,{\partial\over\partial t''}\bigl[\chi_{BA}(t,t'')-\chi_{AB}(t'',t)\bigr]\,{\pi\over\beta\hbar}
\,\coth{\pi(t''-t')\over\beta\hbar}.\eqno(A.9)$$

Like Eq. ($A$.5), Eq. ($A$.9) in the case $A=B$ yields Eq. (2.8) in the classical
limit.

Note that, since Eqs. ($A$.4) and ($A$.8) are equilibrium relations, the various quantities into
play only depend on the time differences involved, so that  Eqs.~($A$.4) and ($A$.8)  are
in fact nothing but convolution products. In the frequency domain and in the case $A=B$, they just
correspond to the usual relations  
$$\chi''(\omega)={1\over\hbar}\,\tanh{\beta\hbar\omega\over 2}\,C(\omega),\qquad
C(\omega)=\hbar\,\coth{\beta\hbar\omega\over 2}\,\chi''(\omega),\eqno(A.9)$$
between the dissipative part $\chi''(\omega)$ of the susceptibility and the Fourier transform
$C(\omega)$ of the correlation function.
\vfill
\break
\parindent=0pt
{\smalltitle References}
\bigskip
\baselineskip=12pt
\frenchspacing

1. See for instance J.-P. Bouchaud, L.F. Cugliandolo, J. Kurchan and M. M\'ezard, in {\sl
Spin-glasses and random fields\/}, A.P. Young Ed. (World Scientific, 1997) and references therein.

2. L.F. Cugliandolo, J. Kurchan and G. Parisi, J. Phys. {\bf 4}, 1641 (1994).

3. W. Krauth and M. M\'ezard, Z. Phys. B {\bf 97}, 127 (1995).

4. F. Ritort, Phys. Rev. Lett. {\bf 75},1190 (1995).

5. C. Godr\`eche and J.M. Luck, J. Phys. A {\bf 29}, 1915 (1995).

6. S. Franz and F. Ritort, Europhys. Lett. {\bf 31}, 507 (1995).

7. L.F. Cugliandolo and G. Lozano, Phys. Rev. Lett. {\bf 80}, 4979 (1998);\break Phys. Rev. B {\bf
59}, 915 (1999).

8. G.M. Sch\"utz and S. Trimper, Europhys. Lett. {\bf 47}, 164 (1999).

9. G.W. Ford, M. Kac and P. Mazur, J. Math. Phys. {\bf 6}, 504 (1965).

10. G.W. Ford and M. Kac, J. Stat. Phys. {\bf 46}, 803 (1987).

11. a) A. Caldeira and A.J. Leggett, Ann. Phys. {\bf 149}, 374 (1983).

b) A.J. Leggett, S. Chakravarty, A.T. Dorsey, M.P.A. Fisher, A. Garg and\break W.Zwerger, Rev. Mod.
Phys. {\bf 59}, 1 (1987).

12. C. Aslangul, N. Pottier and D. Saint-James, J. Stat. Phys. {\bf 40}, 167 (1985).

13. U. Weiss, {\it Quantum dissipative systems\/}, World Scientific, 1993.

14. I.S. Gradshteyn and I.M. Ryzhik, {\it Tables of integrals, series, and products\/}, Academic
Press, 1980.

15. R. M\'elin and P. Butaud, J. Phys. \uppercase\expandafter{\romannumeral 1} {\bf 7}, 691 (1997).

16. C. Aslangul, N. Pottier and D. Saint-James, Phys. Lett. A, {\bf 123}, 413  (1987); J.
Physique {\bf 48}, 1871 (1987).

\vfill
\break
\noindent
{\smalltitle Figure captions}
\bigskip
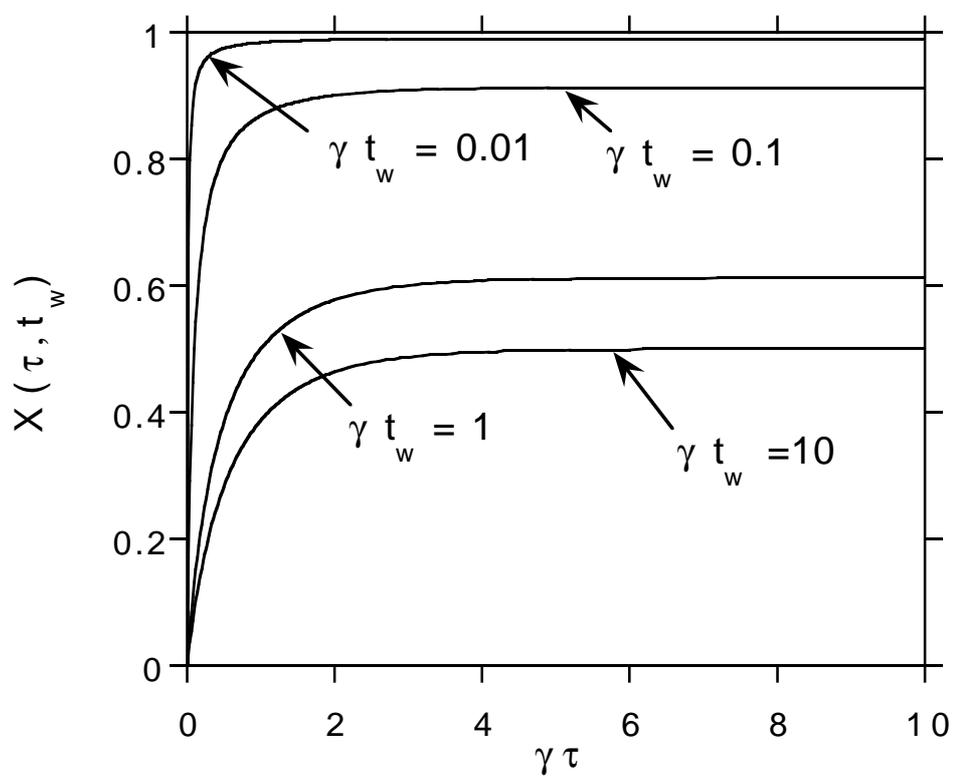
{\bf Fig.~1} 

The violation factor $X(\tau,t_w)={\beta_{\rm eff.}(\tau,t_w)/\beta}$ of the classical 
fluctuation-dissipation theorem in the Langevin model plotted as a function of
$\gamma\tau$ for various values of~$\gamma t_w$: 
$\gamma t_w=0.01$; 
$\gamma t_w=0.1$;
$\gamma t_w=1$; 
$\gamma t_w=10$.
\bigskip
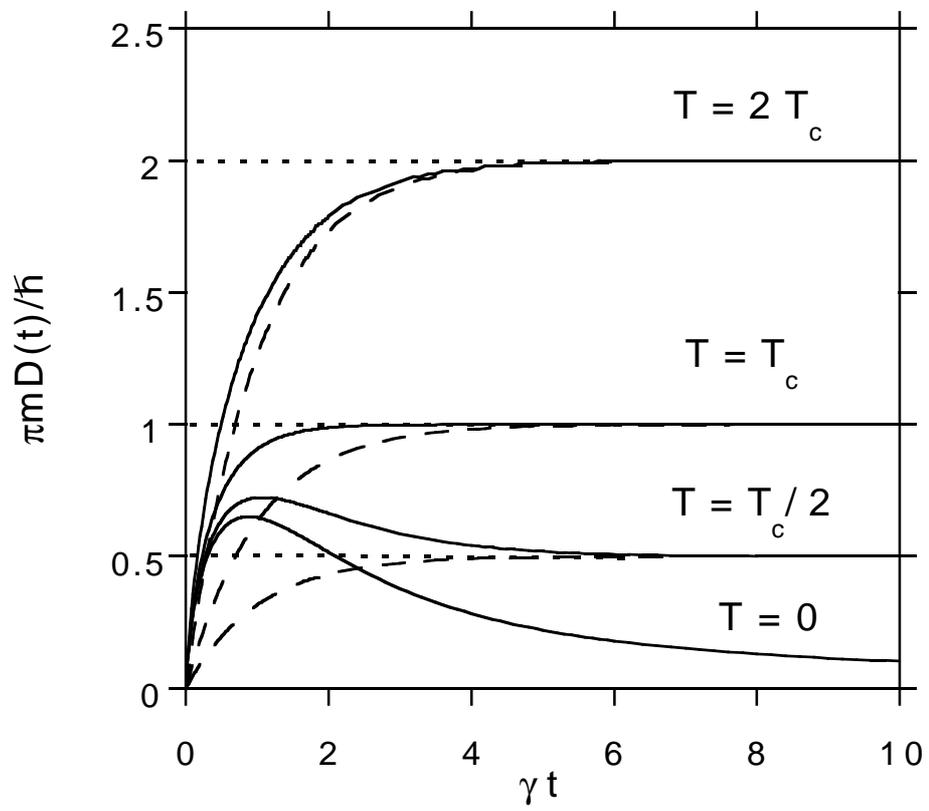
{\bf Fig.~2}

In the short-memory limit ($\omega_c\to\infty$), the quantum time-dependent diffusion
coefficient $D^{Q,\omega_c\to\infty}(t)$ plotted as a function of $\gamma t$ for several
different bath temperatures on both sides of $T_c$ (full lines):

$\gamma t_{\rm th.}=0.25$ ($T=2T_c$, classical regime);\hfill\break
$\gamma t_{\rm th.}=0.5$ ($T=T_c$, cross-over);\hfill\break
$\gamma t_{\rm th.}=1$ ($T={T_c/2}$, quantum regime);\hfill\break
$\gamma t_{\rm th.}=+\infty$ ($T=0$).

The corresponding curves for the classical diffusion coefficient $D^{\omega_c\to\infty}(t)$ are
plotted in dotted lines on the same figure.
\bigskip
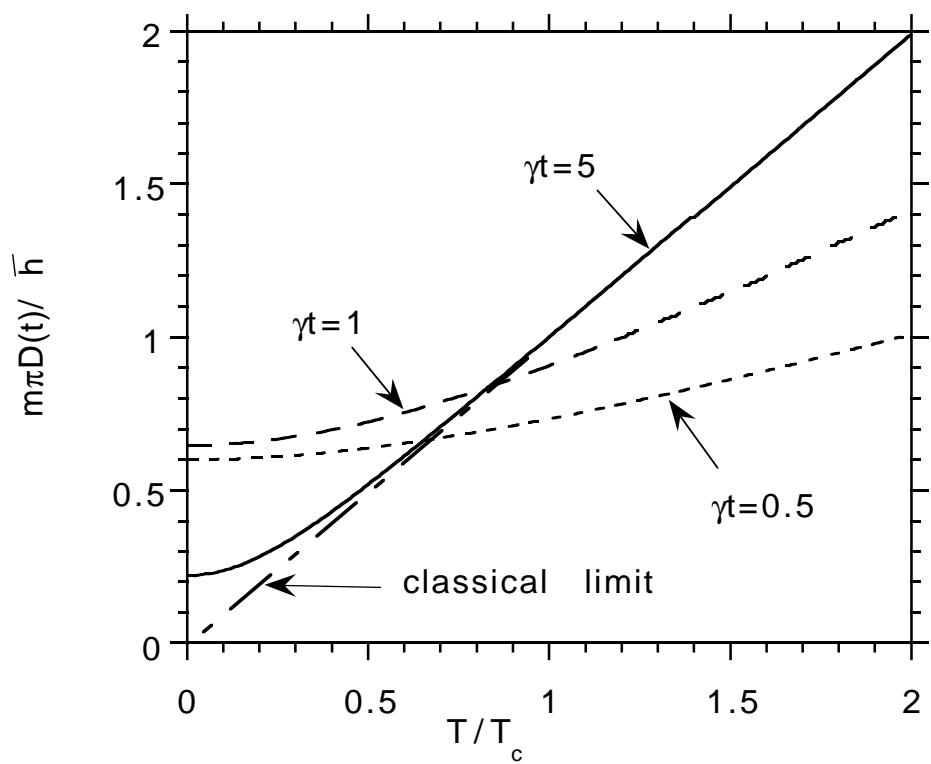
{\bf Fig.~3}

Quantum time-dependent diffusion coefficient $D^{Q,\omega_c\to\infty}(t)$ plotted as a function of
${T/T_c}$ for several different values of $\gamma t$:
$\gamma t=0.5$;
$\gamma t=1$;
$\gamma t=5$;
classical limit ($\gamma t\gg 1$). 
\bigskip
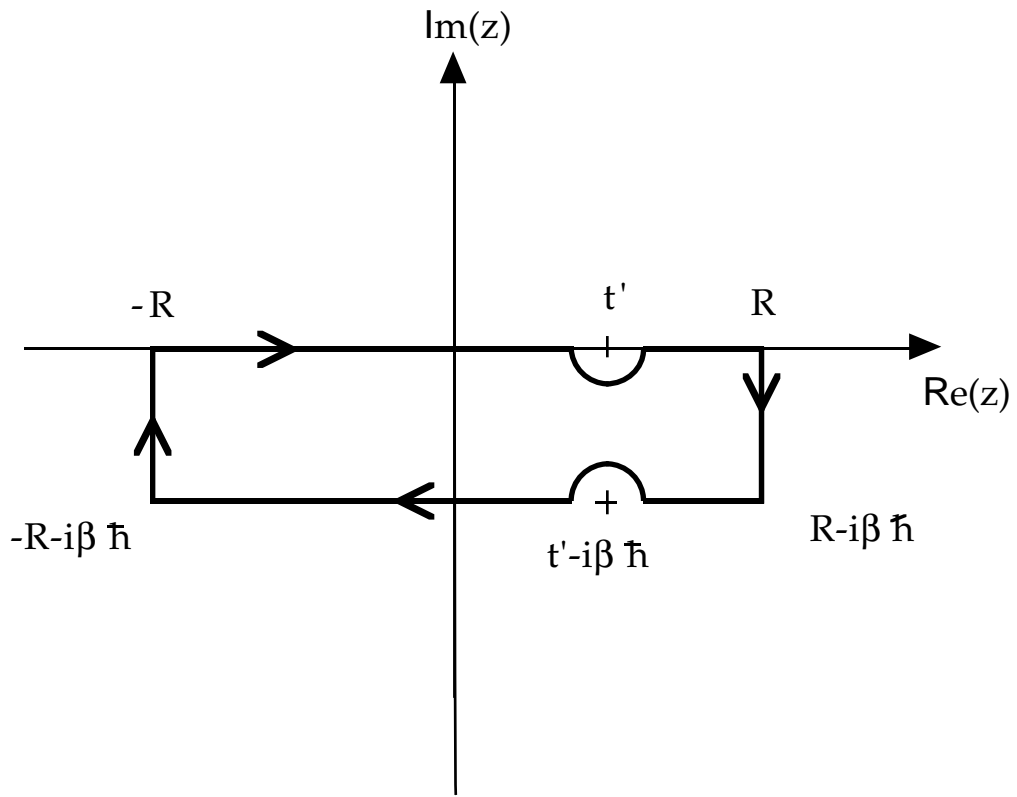
{\bf Fig.~4}

Integration contour for the demonstration of formulas (6.1) and (6.2) of the main text (or formulas
($A$.5) and ($A$.8) of the Appendix).
\bigskip
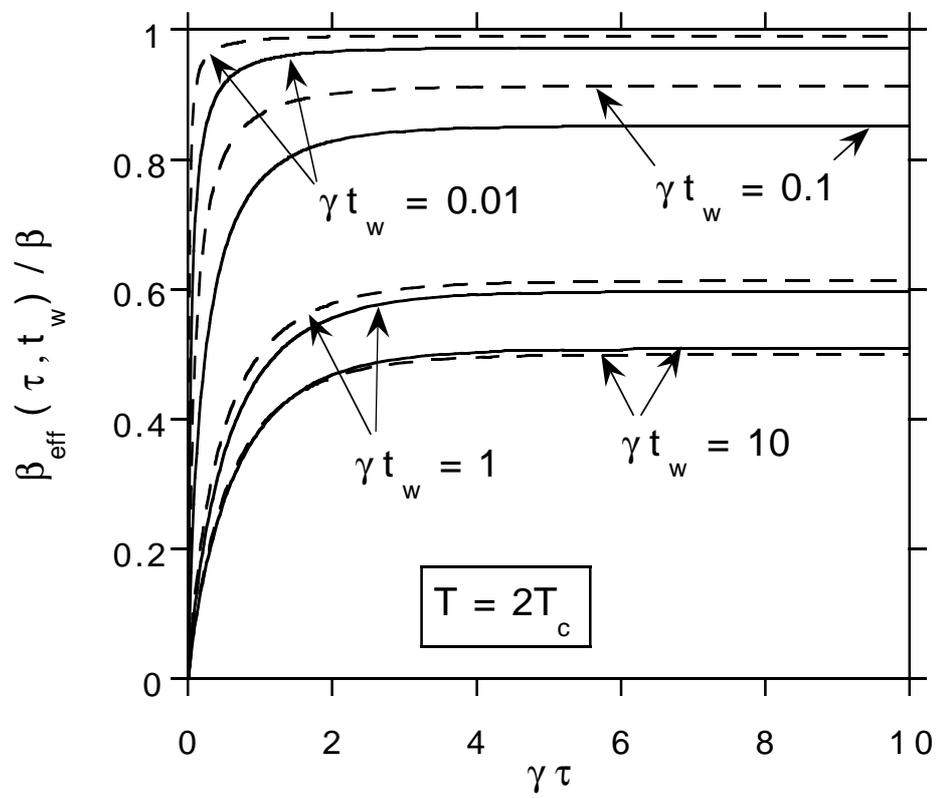
{\bf Fig.~5}

In the short-memory limit ($\omega_c\to\infty$), the violation factor $X^Q(\tau,t_w)={\beta^Q_{\rm
eff.}(\tau,t_w)/\beta}$ of the quantum fluctuation-dissipation theorem, at bath
temperature $T=2T_c$ (classical regime), plotted as a function of
$\gamma\tau$ for various values of $\gamma t_w$ (full lines):
$\gamma t_w=0.01$; 
$\gamma t_w=0.1$;
$\gamma t_w=1$; 
$\gamma t_w=10$.
The corresponding curves for the classical violation factor $X(\tau,t_w)$ are plotted in dotted
lines on the same figure.
\bigskip
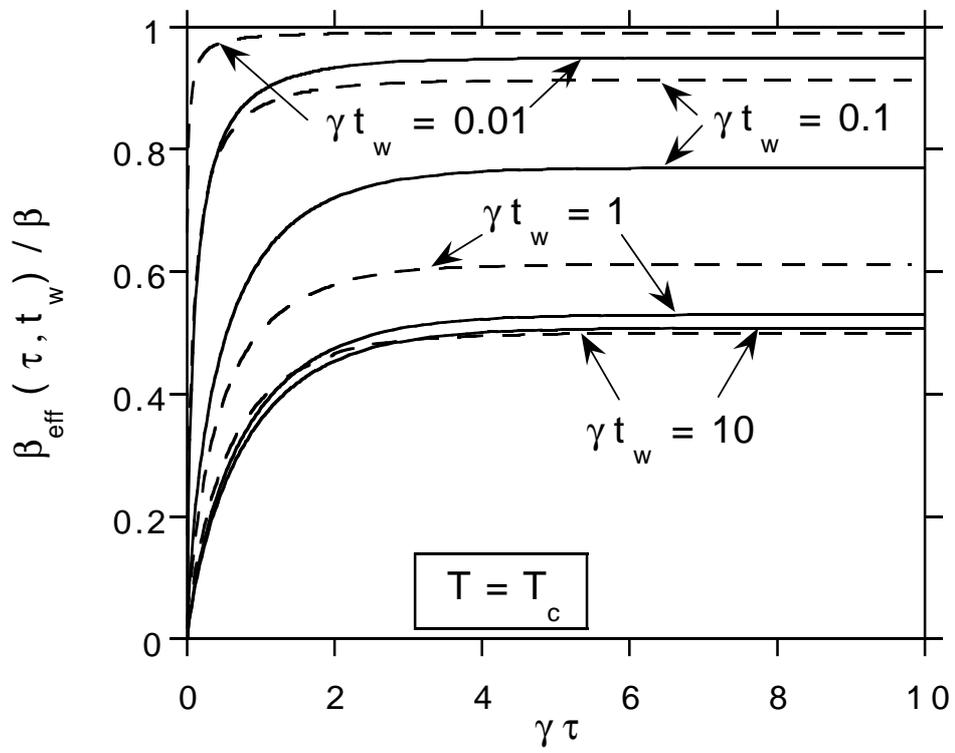
{\bf Fig.~6}

Same as above (Fig.~5), but at bath temperature $T=T_c$ (cross-over).
\bigskip
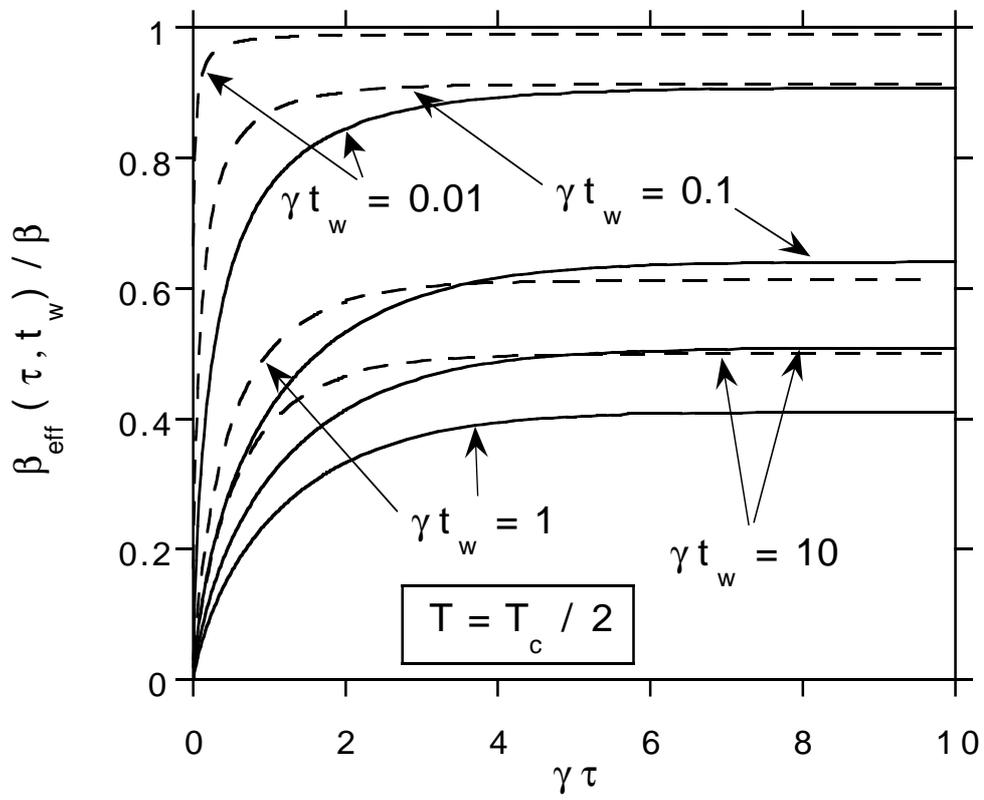
{\bf Fig.~7}

Same as above (Fig.~5 or Fig.~6), but at bath temperature $T={T_c/2}$ (quantum regime).
\bigskip
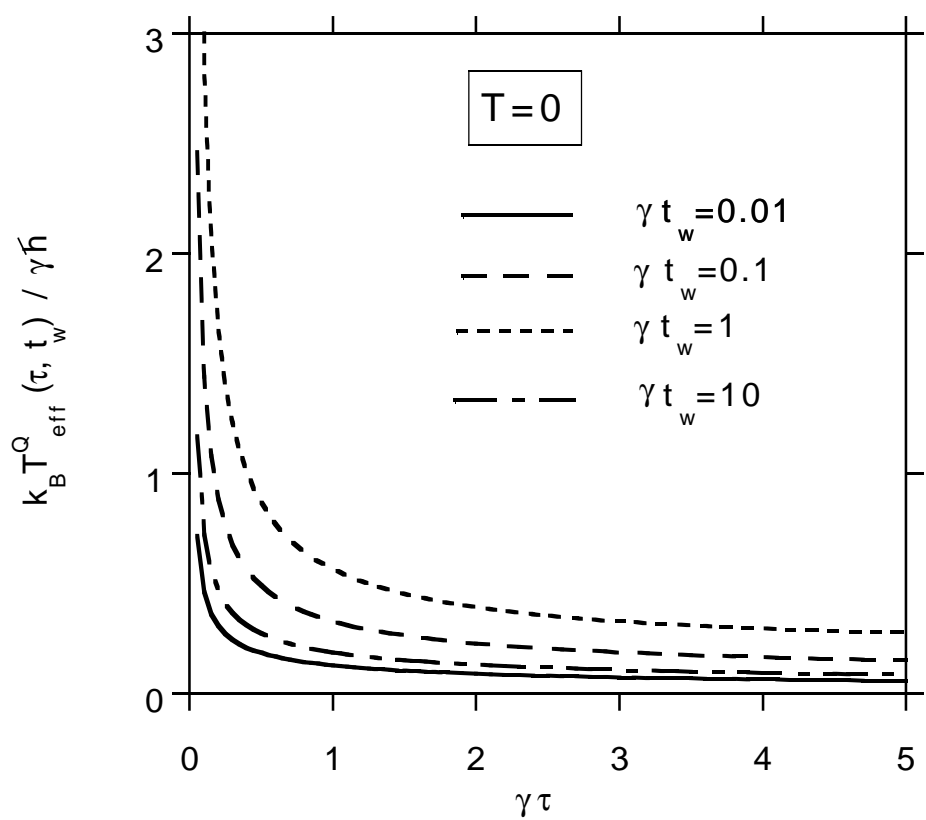
{\bf Fig.~8}

Zero temperature case: the effective temperature $T^Q_{{\rm eff.}}(\tau,t_w)$ plotted as a
function of $\gamma\tau$ for various values of $\gamma t_w$:
$\gamma t_w=0.01$; 
$\gamma t_w=0.1$;
$\gamma t_w=1$; 
$\gamma t_w=10$.
\bigskip
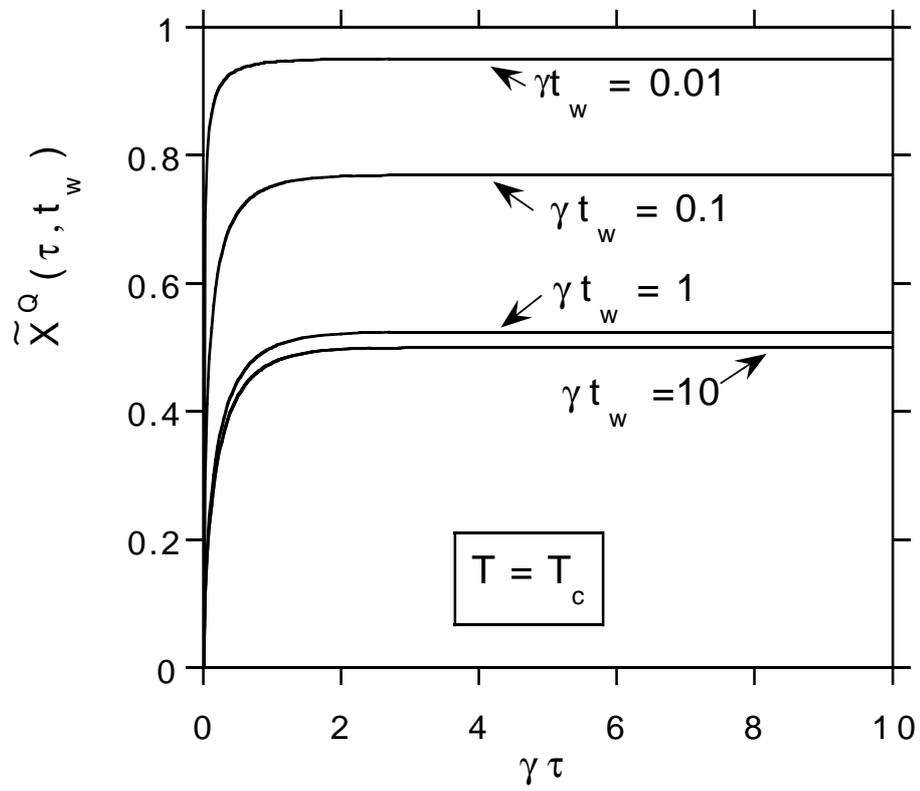
{\bf Fig.~9}

The ratio $\tilde X^Q(\tau,t_w)=D^Q(\tau)/[D^Q(\tau)+D^Q(t_w)]$, at bath temperature $T=T_c$
(cross-over), plotted as a function of $\gamma\tau$ for various values of $\gamma t_w$:
$\gamma t_w=0.01$; 
$\gamma t_w=0.1$;
$\gamma t_w=1$; 
$\gamma t_w=10$.
\bigskip
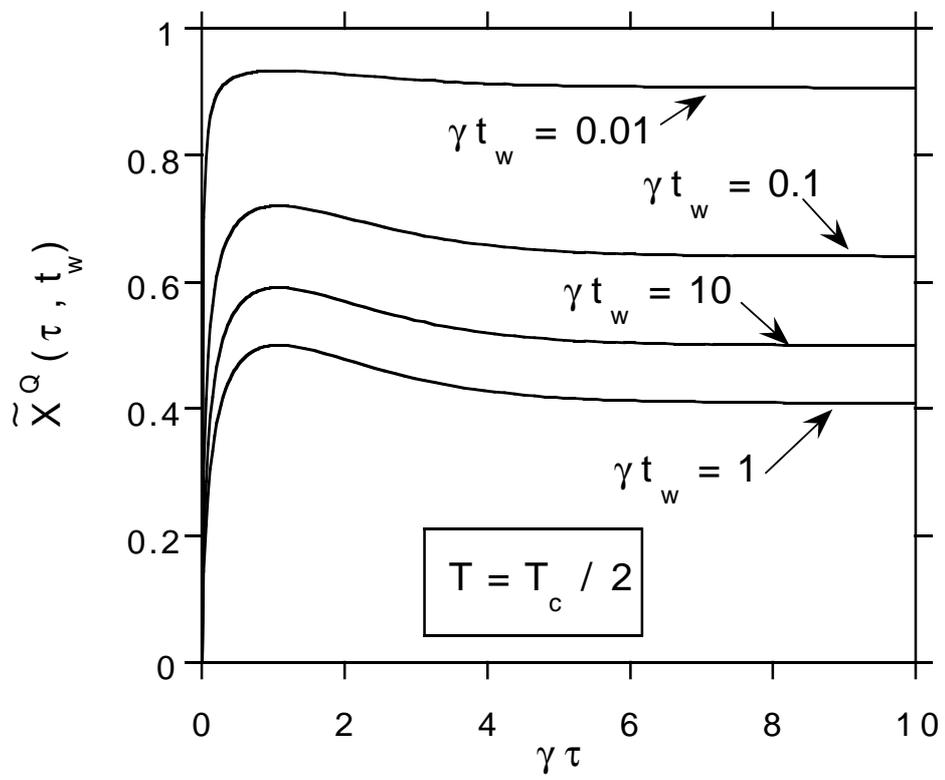
{\bf Fig.~10}

Same as above (Fig.~9), but at bath temperature $T={T_c/2}$ (quantum regime).
\bigskip
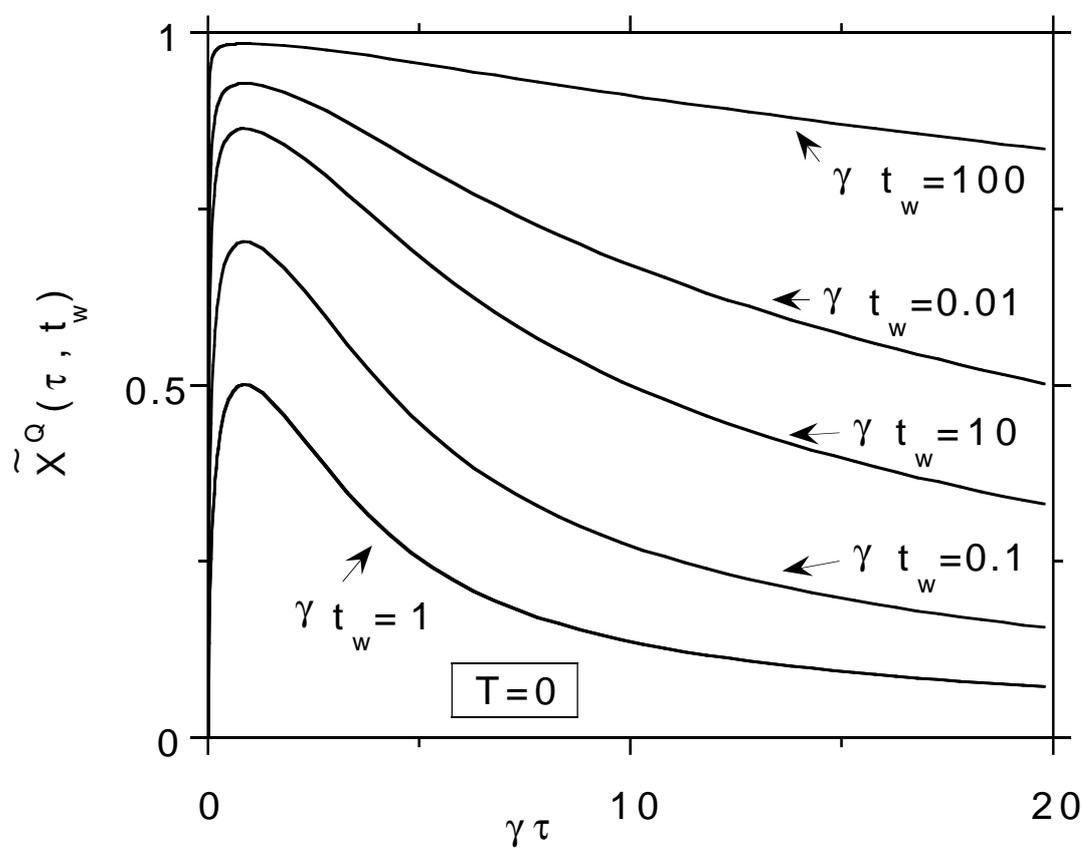
{\bf Fig.~11}

Zero temperature case: $\tilde X^Q(\tau,t_w)$ plotted as a function of $\gamma\tau$ 
for various values of $\gamma t_w$:
$\gamma t_w=0.01$; 
$\gamma t_w=0.1$;
$\gamma t_w=1$; 
$\gamma t_w=10$;
$\gamma t_w=100$.
\bye